\documentclass[twocolumn,letterpaper,aps,prc,longbibliography,superscriptaddress,floatfix]{revtex4-2}

\usepackage{graphicx}
\usepackage{tabularx}
\usepackage{multirow}
\usepackage{amsmath}
\usepackage{booktabs}

\usepackage[dvipsnames]{xcolor}
\usepackage[normalem]{ulem} 
\usepackage{xspace}	

\newcommand{\pt}{\mbox{$p_T$}\xspace}

\newcommand{\Gevc}{\mbox{GeV/$c$}\xspace}
\newcommand{\Mevcc}{\mbox{MeV/c$^2$}\xspace}
\newcommand{\rab}{\mbox{$R_{AB}$}\xspace}
\newcommand{\Npart}{\mbox{$\langle N_{\rm part} \rangle$}\xspace}
\newcommand{\Ncoll}{\mbox{$\langle N_{\rm coll} \rangle$}\xspace}
\newcommand{\kEt}{\mbox{$ KE_T$}\xspace}
\newcommand{\sqsn}{\mbox{$\sqrt{s_{_{NN}}}$}\xspace}
\newcommand{\sqsntwo}{\mbox{$\sqrt{s_{_{NN}}}=200$~GeV}\xspace}
\newcommand{\sqsnuu}{\mbox{$\sqrt{s_{_{NN}}}=193$~GeV}\xspace}
\newcommand{\pp}{\mbox{$p$$+$$p$}\xspace}
\newcommand{\auau}{\mbox{Au$+$Au}\xspace}
\newcommand{\cucu}{\mbox{Cu$+$Cu}\xspace}
\newcommand{\cuau}{\mbox{Cu$+$Au}\xspace}
\newcommand{\uu}{\mbox{U$+$U}\xspace}
\newcommand{\vphi}{\mbox{$\phi$}\xspace}
\newcommand{\pio}{\mbox{$\pi^0$}\xspace}

\newcommand{\et}{\mbox{$\eta$}\xspace}
\newcommand{\prots}{\mbox{$(p+\bar{p})/2$}\xspace}

\newcommand{\vtwo}{\mbox{$v_{2}$}\xspace}
\newcommand{\nq}{\mbox{$n_{q}$}\xspace}
\newcommand{\ecc}{\mbox{$\varepsilon_{2}$}\xspace}
\newcommand{\midrap}{\mbox{$|\eta|<0.35$}\xspace}

\begin{document}

\title{Measurement of $\phi$-meson production in Cu$+$Au collisions at 
$\sqrt{s_{_{NN}}}$ = 200 GeV and U$+$U collisions at $\sqrt{s_{_{NN}}}$ 
= 193 GeV}


\newcommand{\abilene}{Abilene Christian University, Abilene, Texas 79699, USA}
\newcommand{\augie}{Department of Physics, Augustana University, Sioux Falls, South Dakota 57197, USA}
\newcommand{\banaras}{Department of Physics, Banaras Hindu University, Varanasi 221005, India}
\newcommand{\barc}{Bhabha Atomic Research Centre, Bombay 400 085, India}
\newcommand{\baruch}{Baruch College, City University of New York, New York, New York, 10010 USA}
\newcommand{\bnlcoll}{Collider-Accelerator Department, Brookhaven National Laboratory, Upton, New York 11973-5000, USA}
\newcommand{\bnlphys}{Physics Department, Brookhaven National Laboratory, Upton, New York 11973-5000, USA}
\newcommand{\caucr}{University of California-Riverside, Riverside, California 92521, USA}
\newcommand{\charlesczech}{Charles University, Faculty of Mathematics and Physics, 180 00 Troja, Prague, Czech Republic}
\newcommand{\ciae}{Science and Technology on Nuclear Data Laboratory, China Institute of Atomic Energy, Beijing 102413, People's Republic of China}
\newcommand{\cns}{Center for Nuclear Study, Graduate School of Science, University of Tokyo, 7-3-1 Hongo, Bunkyo, Tokyo 113-0033, Japan}
\newcommand{\colorado}{University of Colorado, Boulder, Colorado 80309, USA}
\newcommand{\columbia}{Columbia University, New York, New York 10027 and Nevis Laboratories, Irvington, New York 10533, USA}
\newcommand{\czechtech}{Czech Technical University, Zikova 4, 166 36 Prague 6, Czech Republic}
\newcommand{\debrecen}{Debrecen University, H-4010 Debrecen, Egyetem t{\'e}r 1, Hungary}
\newcommand{\elte}{ELTE, E{\"o}tv{\"o}s Lor{\'a}nd University, H-1117 Budapest, P{\'a}zm{\'a}ny P.~s.~1/A, Hungary}
\newcommand{\ewha}{Ewha Womans University, Seoul 120-750, Korea}
\newcommand{\famu}{Florida A\&M University, Tallahassee, FL 32307, USA}
\newcommand{\fsu}{Florida State University, Tallahassee, Florida 32306, USA}
\newcommand{\gsu}{Georgia State University, Atlanta, Georgia 30303, USA}
\newcommand{\hanyang}{Hanyang University, Seoul 133-792, Korea}
\newcommand{\hiroshima}{Hiroshima University, Kagamiyama, Higashi-Hiroshima 739-8526, Japan}
\newcommand{\howard}{Department of Physics and Astronomy, Howard University, Washington, DC 20059, USA}
\newcommand{\ihepprot}{IHEP Protvino, State Research Center of Russian Federation, Institute for High Energy Physics, Protvino, 142281, Russia}
\newcommand{\illuiuc}{University of Illinois at Urbana-Champaign, Urbana, Illinois 61801, USA}
\newcommand{\inrras}{Institute for Nuclear Research of the Russian Academy of Sciences, prospekt 60-letiya Oktyabrya 7a, Moscow 117312, Russia}
\newcommand{\instpasczech}{Institute of Physics, Academy of Sciences of the Czech Republic, Na Slovance 2, 182 21 Prague 8, Czech Republic}
\newcommand{\isu}{Iowa State University, Ames, Iowa 50011, USA}
\newcommand{\jaea}{Advanced Science Research Center, Japan Atomic Energy Agency, 2-4 Shirakata Shirane, Tokai-mura, Naka-gun, Ibaraki-ken 319-1195, Japan}
\newcommand{\jeonbuk}{Jeonbuk National University, Jeonju, 54896, Korea}
\newcommand{\jyvaskyla}{Helsinki Institute of Physics and University of Jyv{\"a}skyl{\"a}, P.O.Box 35, FI-40014 Jyv{\"a}skyl{\"a}, Finland}
\newcommand{\kek}{KEK, High Energy Accelerator Research Organization, Tsukuba, Ibaraki 305-0801, Japan}
\newcommand{\korea}{Korea University, Seoul 02841, Korea}
\newcommand{\kurchatov}{National Research Center ``Kurchatov Institute", Moscow, 123098 Russia}
\newcommand{\kyoto}{Kyoto University, Kyoto 606-8502, Japan}
\newcommand{\labllr}{Laboratoire Leprince-Ringuet, Ecole Polytechnique, CNRS-IN2P3, Route de Saclay, F-91128, Palaiseau, France}
\newcommand{\lahorelums}{Physics Department, Lahore University of Management Sciences, Lahore 54792, Pakistan}
\newcommand{\lawllnl}{Lawrence Livermore National Laboratory, Livermore, California 94550, USA}
\newcommand{\losalamos}{Los Alamos National Laboratory, Los Alamos, New Mexico 87545, USA}
\newcommand{\lund}{Department of Physics, Lund University, Box 118, SE-221 00 Lund, Sweden}
\newcommand{\maryland}{University of Maryland, College Park, Maryland 20742, USA}
\newcommand{\mass}{Department of Physics, University of Massachusetts, Amherst, Massachusetts 01003-9337, USA}
\newcommand{\mate}{MATE, Laboratory of Femtoscopy, K\'aroly R\'obert Campus, H-3200 Gy\"ongy\"os, M\'atrai\'ut 36, Hungary}
\newcommand{\michigan}{Department of Physics, University of Michigan, Ann Arbor, Michigan 48109-1040, USA}
\newcommand{\miss}{Mississippi State University, Mississippi State, Mississippi 39762, USA}
\newcommand{\muhlenberg}{Muhlenberg College, Allentown, Pennsylvania 18104-5586, USA}
\newcommand{\myongji}{Myongji University, Yongin, Kyonggido 449-728, Korea}
\newcommand{\nagasaki}{Nagasaki Institute of Applied Science, Nagasaki-shi, Nagasaki 851-0193, Japan}
\newcommand{\nara}{Nara Women's University, Kita-uoya Nishi-machi Nara 630-8506, Japan}
\newcommand{\natmephi}{National Research Nuclear University, MEPhI, Moscow Engineering Physics Institute, Moscow, 115409, Russia}
\newcommand{\newmex}{University of New Mexico, Albuquerque, New Mexico 87131, USA}
\newcommand{\nmsu}{New Mexico State University, Las Cruces, New Mexico 88003, USA}
\newcommand{\northcg}{Physics and Astronomy Department, University of North Carolina at Greensboro, Greensboro, North Carolina 27412, USA}
\newcommand{\ohio}{Department of Physics and Astronomy, Ohio University, Athens, Ohio 45701, USA}
\newcommand{\ornl}{Oak Ridge National Laboratory, Oak Ridge, Tennessee 37831, USA}
\newcommand{\orsay}{IPN-Orsay, Univ.~Paris-Sud, CNRS/IN2P3, Universit\'e Paris-Saclay, BP1, F-91406, Orsay, France}
\newcommand{\pnpi}{PNPI, Petersburg Nuclear Physics Institute, Gatchina, Leningrad region, 188300, Russia}
\newcommand{\pusan}{Pusan National University, Pusan 46241, Korea}
\newcommand{\riken}{RIKEN Nishina Center for Accelerator-Based Science, Wako, Saitama 351-0198, Japan}
\newcommand{\rikjrbrc}{RIKEN BNL Research Center, Brookhaven National Laboratory, Upton, New York 11973-5000, USA}
\newcommand{\rikkyo}{Physics Department, Rikkyo University, 3-34-1 Nishi-Ikebukuro, Toshima, Tokyo 171-8501, Japan}
\newcommand{\saispbstu}{Saint Petersburg State Polytechnic University, St.~Petersburg, 195251 Russia}
\newcommand{\seoulnat}{Department of Physics and Astronomy, Seoul National University, Seoul 151-742, Korea}
\newcommand{\stonybrkc}{Chemistry Department, Stony Brook University, SUNY, Stony Brook, New York 11794-3400, USA}
\newcommand{\stonycrkp}{Department of Physics and Astronomy, Stony Brook University, SUNY, Stony Brook, New York 11794-3800, USA}
\newcommand{\sungskku}{Sungkyunkwan University, Suwon, 440-746, Korea}
\newcommand{\tenn}{University of Tennessee, Knoxville, Tennessee 37996, USA}
\newcommand{\texsu}{Texas Southern University, Houston, TX 77004, USA}
\newcommand{\titech}{Department of Physics, Tokyo Institute of Technology, Oh-okayama, Meguro, Tokyo 152-8551, Japan}
\newcommand{\tsukuba}{Tomonaga Center for the History of the Universe, University of Tsukuba, Tsukuba, Ibaraki 305, Japan}
\newcommand{\vandy}{Vanderbilt University, Nashville, Tennessee 37235, USA}
\newcommand{\weizmann}{Weizmann Institute, Rehovot 76100, Israel}
\newcommand{\wigner}{Institute for Particle and Nuclear Physics, Wigner Research Centre for Physics, Hungarian Academy of Sciences (Wigner RCP, RMKI) H-1525 Budapest 114, POBox 49, Budapest, Hungary}
\newcommand{\yonsei}{Yonsei University, IPAP, Seoul 120-749, Korea}
\newcommand{\zagreb}{Department of Physics, Faculty of Science, University of Zagreb, Bijeni\v{c}ka c.~32 HR-10002 Zagreb, Croatia}
\newcommand{\zambia}{Department of Physics, School of Natural Sciences, University of Zambia, Great East Road Campus, Box 32379, Lusaka, Zambia}
\affiliation{\abilene}
\affiliation{\augie}
\affiliation{\banaras}
\affiliation{\barc}
\affiliation{\baruch}
\affiliation{\bnlcoll}
\affiliation{\bnlphys}
\affiliation{\caucr}
\affiliation{\charlesczech}
\affiliation{\ciae}
\affiliation{\cns}
\affiliation{\colorado}
\affiliation{\columbia}
\affiliation{\czechtech}
\affiliation{\debrecen}
\affiliation{\elte}
\affiliation{\ewha}
\affiliation{\famu}
\affiliation{\fsu}
\affiliation{\gsu}
\affiliation{\hanyang}
\affiliation{\hiroshima}
\affiliation{\howard}
\affiliation{\ihepprot}
\affiliation{\illuiuc}
\affiliation{\inrras}
\affiliation{\instpasczech}
\affiliation{\isu}
\affiliation{\jaea}
\affiliation{\jeonbuk}
\affiliation{\jyvaskyla}
\affiliation{\kek}
\affiliation{\korea}
\affiliation{\kurchatov}
\affiliation{\kyoto}
\affiliation{\labllr}
\affiliation{\lahorelums}
\affiliation{\lawllnl}
\affiliation{\losalamos}
\affiliation{\lund}
\affiliation{\maryland}
\affiliation{\mass}
\affiliation{\mate}
\affiliation{\michigan}
\affiliation{\miss}
\affiliation{\muhlenberg}
\affiliation{\myongji}
\affiliation{\nagasaki}
\affiliation{\nara}
\affiliation{\natmephi}
\affiliation{\newmex}
\affiliation{\nmsu}
\affiliation{\northcg}
\affiliation{\ohio}
\affiliation{\ornl}
\affiliation{\orsay}
\affiliation{\pnpi}
\affiliation{\pusan}
\affiliation{\riken}
\affiliation{\rikjrbrc}
\affiliation{\rikkyo}
\affiliation{\saispbstu}
\affiliation{\seoulnat}
\affiliation{\stonybrkc}
\affiliation{\stonycrkp}
\affiliation{\sungskku}
\affiliation{\tenn}
\affiliation{\texsu}
\affiliation{\titech}
\affiliation{\tsukuba}
\affiliation{\vandy}
\affiliation{\weizmann}
\affiliation{\wigner}
\affiliation{\yonsei}
\affiliation{\zagreb}
\affiliation{\zambia}
\author{N.J.~Abdulameer} \affiliation{\debrecen}
\author{U.~Acharya} \affiliation{\gsu} 
\author{C.~Aidala} \affiliation{\losalamos} \affiliation{\michigan} 
\author{N.N.~Ajitanand} \altaffiliation{Deceased} \affiliation{\stonybrkc} 
\author{Y.~Akiba} \email[PHENIX Spokesperson: ]{akiba@rcf.rhic.bnl.gov} \affiliation{\riken} \affiliation{\rikjrbrc} 
\author{R.~Akimoto} \affiliation{\cns} 
\author{J.~Alexander} \affiliation{\stonybrkc} 
\author{M.~Alfred} \affiliation{\howard} 
\author{M.~Alibordi} \affiliation{\miss}
\author{K.~Aoki} \affiliation{\kek} \affiliation{\riken} 
% ...
\collaboration{PHENIX Collaboration}  \noaffiliation

\date{\today}



\begin{abstract}


The PHENIX experiment reports systematic measurements at the 
Relativistic Heavy Ion Collider of $\phi$-meson production in asymmetric 
Cu$+$Au collisions at $\sqrt{s_{_{NN}}}$~=~200 GeV and in U$+$U collisions 
at $\sqrt{s_{_{NN}}}$~=~193 GeV.  Measurements were performed via the 
$\phi\rightarrow K^{+}K^{-}$ decay channel at midrapidity $|\eta|<0.35$.  
Features of $\phi$-meson production measured in Cu$+$Cu, Cu$+$Au, 
Au$+$Au, and U$+$U collisions were found to not depend on the collision 
geometry, which was expected because the yields are averaged over the 
azimuthal angle and follow the expected scaling with nuclear-overlap 
size. The elliptic flow of the $\phi$ meson in Cu$+$Au, Au$+$Au, and 
U$+$U collisions scales with second-order-participant eccentricity and 
the length scale of the nuclear-overlap region (estimated with the 
number of participating nucleons). At moderate $p_T$, $\phi$-meson 
production measured in Cu$+$Au and U$+$U collisions is consistent with 
coalescence-model predictions, whereas at high $p_T$ the production is 
in agreement with expectations for in-medium energy loss of parent 
partons prior to their fragmentation.  The elliptic flow for $\phi$ 
mesons measured in Cu$+$Au and U$+$U collisions is well described by a 
(2+1)D viscous-hydrodynamic model with specific-shear viscosity 
$\eta/s=1/4\pi$.

\end{abstract}


\maketitle

\section{INTRODUCTION}

The formation of the quark-gluon plasma (QGP) has been established by 
experiments at the Relativistic Heavy Ion Collider 
(RHIC)~\cite{RHIC,QGP,BRAHMS,PHOBOS,STAR} and later at the Large Hadron 
Collider (LHC)~\cite{LHC,ALICE,ATLAS,ALICE1}. Since then, one of the 
main goals of high-energy nuclear physics, including the PHENIX 
experiment~\cite{PHENIXoverview}, is to quantify and characterize the 
properties of the QGP. Measurements of light-hadron production in 
collision systems with different geometries are commonly used for the 
systematic experimental study of the evolution of the medium created in 
high-energy nuclear collisions, including the QGP phase.

The processes of QGP formation and evolution depend on the initial 
conditions. These include the collision-system energy, 
the nuclear-overlap size and shape, and nuclear modification of the 
parton-distribution functions~\cite{nPDFs}. In the most central \cuau 
collisions, the Cu ion is fully occluded by the Au ion, which might lead 
to significantly larger suppression of particle yields than for 
symmetric systems like \cucu and \auau~\cite{Pi_eta_cuau}. Collisions of 
uranium nuclei, which are highly deformed, provide different collision 
configurations depending on their orientation relative to the reaction 
plane. On average, comparing to symmetric systems, the nuclear-overlap 
region in \cuau and \uu collisions has additional asymmetry along the 
impact-parameter orientation. In addition to different nuclear 
thicknesses, this leads to initial conditions that are different from 
those of symmetric systems.  Thus, comparison of particle production 
measured in \cuau and \uu collision and symmetric systems is a useful 
tool to study the influence of initial conditions on the evolution of 
heavy ion collisions.

Measurements of light-hadron transverse-momentum (\pt) spectra provide 
probes of QGP effects, such as jet quenching~\cite{JetQuenching} and 
strangeness enhancement~\cite{StrangenessEnhancenment}. Jet quenching 
manifests as a suppression of high-\pt hadron yields, due to parton 
energy losses in the hot and dense medium. Strangeness enhancement can 
be observed as the increase of strange and hidden-strange hadron yields 
in nucleus-nucleus collisions relative to \pp collision scaled by the 
appropriate number of binary nucleon-nucleon collisions. Hadronization 
via the parton-coalescence (recombination) mechanism~\cite{Recombination1, 
Recombination2, Recombination_phi} should be considered to quantify 
strangeness-enhancement effect. Because the degree of strangeness saturation 
and parton energy loss are sensitive to initial 
conditions~\cite{JetQuenching,StrangenessEnhancenment}, measurements of 
strange and hidden-strange hadron production can shed light on the 
physics of the initial conditions.

The investigation of elliptic-flow coefficients (\vtwo) can provide 
insight on how the initial transverse coordinate-space anisotropy of 
heavy ion collisions is converted to a momentum-space anisotropy in the 
transverse plane~\cite{Flow_meth}. Previous studies of \vtwo at RHIC in 
symmetric collision systems (\auau, \cucu) show that \vtwo values for 
light hadrons depend on the number of valence quarks in the hadron 
(\nq), the second-order-participant eccentricity (\ecc) and the number 
of nucleons participating in the interaction 
(\Npart)~\cite{FlowCuCuAuAu,Flow}. The comparison of obtained results to 
the hydrodynamic model predictions suggests that the QGP has properties 
of a nearly perfect fluid~\cite{hydrodynamic}. Therefore, hadronic 
elliptic flow is sensitive to the shape and the size of the 
nuclear-overlap region. Measurements of \vtwo for light hadrons in \cuau 
and \uu collisions can reveal underlying physics mechanisms of its 
development.

The \vphi meson is considered to be a clean probe of QGP properties. It 
has an Okubo-Zweig-Iizuka (OZI) suppressed-interaction cross section 
with nonstrange hadrons and a lifetime (42 fm/c~\cite{PhiQGP}) longer 
than that of the fireball before freeze-out ($\approx$10-20 
fm/c~\cite{QGP_expansion_rates,AMPT, iebevishnu}). Therefore, \vphi 
mesons mostly decay outside the fireball, and along with their daughter 
particles rescatter less frequently in the post-hadronization phase. 
Consequently, the kinematic properties are primarily controlled by 
conditions in the early partonic phase and less affected in the 
hadronization stage. The \vphi vector meson is the nearly pure lightest 
bound state of $s\bar{s}$ quarks. Accordingly, measurements of the 
\vphi-meson \pt spectra in various collision systems can contribute to 
the understanding of strangeness enhancement, along with energy loss and 
coalescence. The comparison of \vtwo values for \vphi mesons to nonOZI 
suppressed $\pi^\pm$ mesons and (anti)protons can indicate a role of 
hadronization stage in \vtwo development.

The PHENIX experiment has measured \vphi-meson production in asymmetric 
\cuau collisions at \sqsntwo and in the largest collision system at RHIC 
\uu at \sqsnuu. The influence of initial conditions on \vphi-meson 
production is investigated by measuring invariant \pt spectra, \rab, and 
\vtwo in \cuau and \uu collisions. The obtained results are compared to 
theoretical calculations based on viscous hydrodynamics 
(iEBE-VISHNU~\cite{iebevishnu}), a-multiphase-transport model 
(AMPT~\cite{AMPT}) and a leading-order (LO) perturbative quantum 
chromodynamics (pQCD) model (PYTHIA/Angantyr~\cite{PYTHIA_ANGANTYR}).

\section{DATA ANALYSIS}

\subsection{Data sets and event selection}

\begin{figure}[htb]
\includegraphics[width=0.96\linewidth]{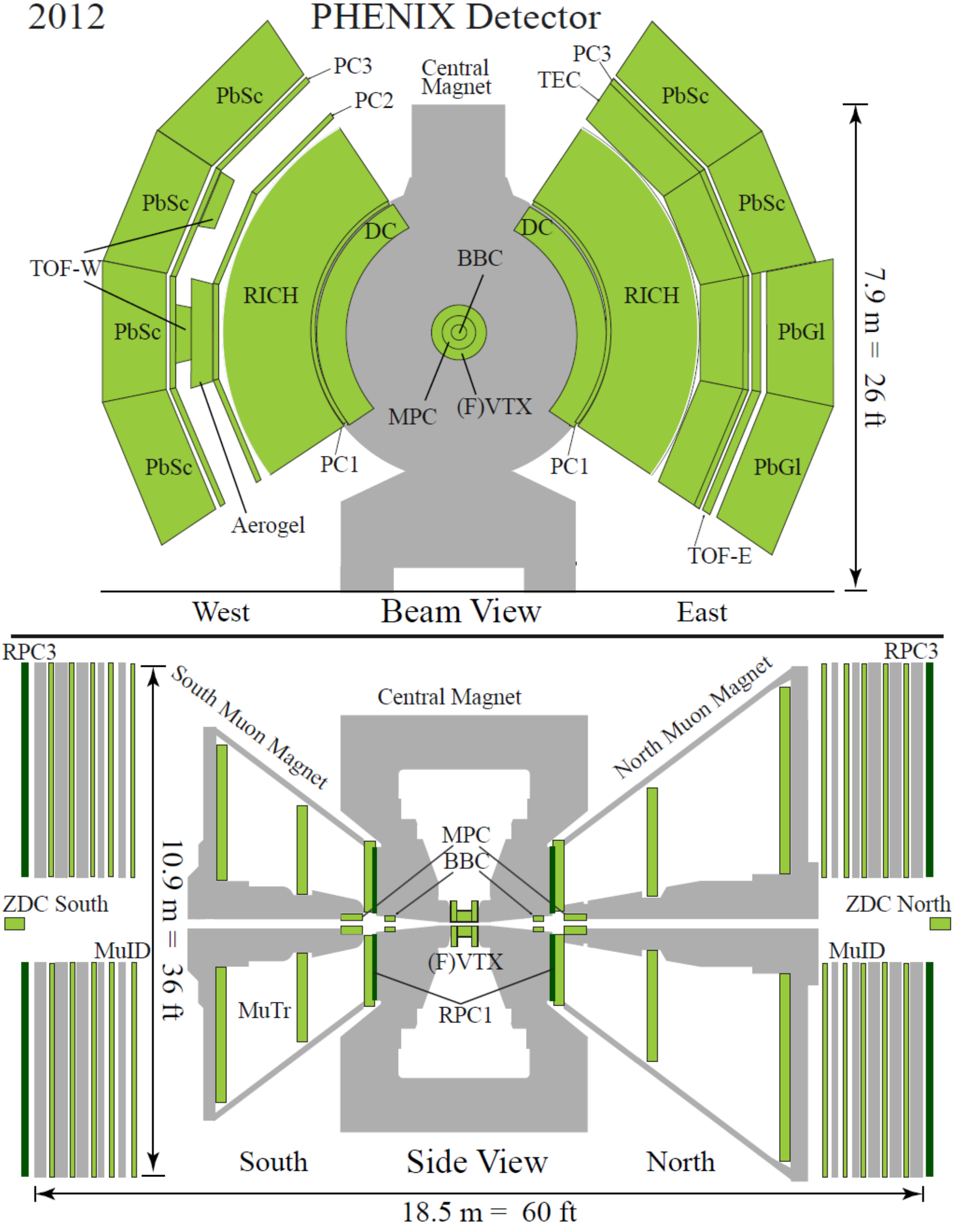}
\caption{The PHENIX detector configuration for data taking in 2012.}
\label{fig:PHENIX}
\end{figure}

The \vphi-meson production analyses are based on data sets collected 
from \cuau collisions at \sqsntwo and \uu collisions at \sqsnuu by the 
PHENIX detector during the 2012 running period. Figure~\ref{fig:PHENIX}
shows the relevant experimental 
setup~\cite{PHENIXoverview}.

The PHENIX detector has two beam-beam counters (BBC)~\cite{BBC} located 
at $\pm144$ cm from the nominal interaction point, each of which covers 
$0 < \phi < 2\pi$ in azimuthal angle and $3.1 < |\eta| < 3.9$ in 
pseudorapidity. The minimum-bias (MB) trigger requires at least two 
phototubes on each side of the BBC to have a signal above the noise 
threshold. The MB definition is satisfied by $93 \pm 2\%$ of the 
inelastic \cuau and \uu cross section. The online $z$-vertex of the 
event is determined by the time difference between signals from the north 
and south arms of the BBC, and is required to be within $\pm30$ cm from 
the center of the detector. The recorded luminosity of \cuau and \uu 
collisions is 27.0 ${\rm nb}^{-1}$ and 736 $\mu{\rm b}^{-1}$, 
respectively.

\subsection{Centrality and event plane azimuthal angle}

The event centrality class in \cuau and \uu collisions is determined as 
a percentile of the absolute values of the total charge measured in the north 
and south BBCs~\cite{Centrality}. Glauber model Monte-Carlo 
simulations~\cite{Glauber} that include the responses of the BBC are 
used to estimate the average values of the number of participating 
nucleons \Npart, the number of binary nucleon-nucleon collisions \Ncoll, 
and the second-harmonic eccentricity \ecc for each centrality class. Two 
parameterizations of the deformed Woods-Saxon distribution for Uranium 
nuclei are considered because there is no single universally accepted 
parameterization of the U nucleus. Two Monte-Carlo simulations were 
produced to provide two sets, Glauber 1~\cite{Glauber1} and Glauber 
2~\cite{Glauber2}, of the collision geometry parameters. The \Npart, 
\Ncoll, and \ecc values for \cuau and \uu collisions are presented in 
Tables~\ref{table:Ncoll_Npart_Cuau_UU}-\ref{table:ecc}.


\begin{table}[tbh]
\caption{\label{table:Ncoll_Npart_Cuau_UU}
Values of \Ncoll and  \Npart for MB and centrality ranges in
\cuau collisions at \sqsn = 200 GeV and \uu collisions at \sqsn = 193 GeV }.
\begin{ruledtabular}
\begin{tabular}{ccccc}
Coll. & Glauber & Centrality   & \Ncoll        & \Npart           \\ \hline
\cuau & Glauber         & MB & 108     $\pm$ 11  &  61.1 $\pm$ 2.7    \\
      & Ref.~\cite{Glauber}  & 0\%--20\%       & 313  $\pm$ 28  &   154 $\pm$ 4  \\
      &      & 20\%--40\% & 129  $\pm$ 12  &  80.4 $\pm$ 3.3    \\
      &      & 40\%--60\% & 41.8 $\pm$ 5.3 &  34.9 $\pm$ 2.8    \\
      &      & 20\%--60\% & 85.6 $\pm$ 8.9 &  57.7 $\pm$ 3.1    \\
      &      & 60\%--80\% & 10.1 $\pm$ 2.0 &  12.1 $\pm$ 1.9    \\
\uu   & Glauber 1           & 0\%--80\%     & 342 $\pm$ 30    & 143 $\pm$ 5      \\
      & Ref.~\cite{Glauber1}& 0\%--20\%     & 935 $\pm$ 98    & 330 $\pm$ 6      \\
      &                     & 20\%--40\%    & 335 $\pm$ 33    & 159 $\pm$ 7      \\
      &                     & 40\%--60\%    & 81.0 $\pm$ 13.1 & 64.8 $\pm$ 5.9   \\
      &                     & 60\%--80\%    & 17.5 $\pm$ 3.9  & 17.8 $\pm$ 3.2   \\
\uu   & Glauber 2           & 0\%--80\%     & 375 $\pm$ 42    & 144 $\pm$ 5      \\
      & Ref.~\cite{Glauber2}& 0\%--20\%     & 999 $\pm$ 114   & 330 $\pm$ 6      \\
      &                     & 20\%--40\%    & 375 $\pm$ 46    & 161 $\pm$ 7      \\
      &                     & 40\%--60\%    & 110 $\pm$ 15    & 65.8 $\pm$ 5.8   \\
      &                     & 60\%--80\%    & 19.8 $\pm$ 4.4  & 18.2 $\pm$ 3.2
\end{tabular} \end{ruledtabular}
\end{table}

\begin{table}[tbh]
\caption{\label{table:ecc}
Values of \ecc for \cuau collisions at \sqsn = 200 GeV and \uu collisions at 
\sqsn = 193 GeV.}
\begin{ruledtabular}
\begin{tabular}{cccc}
Collisions & Glauber & Centrality &\ecc              \\ \hline
\cuau & Glauber~\cite{Glauber}    & 0\%--20\%   & 0.171 $\pm$ 0.009  \\
      &                           & 20\%--40\%  & 0.318 $\pm$ 0.009  \\
      &                           & 40\%--60\%  & 0.480 $\pm$ 0.016  \\
      &                           & 20\%--60\%  & 0.399 $\pm$ 0.012  \\
\uu   & Glauber 1~\cite{Glauber1} & 0\%--50\%   & 0.310 $\pm$ 0.024  \\
\uu   & Glauber 2~\cite{Glauber2} & 0\%--50\%   & 0.366 $\pm$ 0.013  \\
\end{tabular}
\end{ruledtabular}
\end{table}

The azimuthal angle of the event plane $\Psi_2$ is determined using 
the forward silicon-vertex detector (FVTX)~\cite{FVTX} in \cuau collisions 
and the muon-piston calorimeter (MPC)~\cite{MPC} in \uu collisions. The 
$\Psi_2$ obtained with the BBC detector is used to estimate systematic 
uncertainties in both collision systems. The FVTX is a silicon detector 
designed to provide precise tracking for charged particles entering the 
muon spectrometer before undergoing multiple scattering in the hadron 
absorber. The FVTX comprises two arms, north and south, covering a large
pseudorapidity interval $1 < |\eta| < 3$. 
The MPC is a 
lead-tungstate calorimeter equipped with PbWO$_4$ crystal scintillator 
towers. The north arm of the MPC has 220 towers spanning 
pseudorapidities $3.1 < \eta < 3.9$, while the south MPC has 196 towers 
spanning $-3.7 < \eta < -3.1$. The MPC covers almost the same \et range 
as the BBC, but has finer granularity and detects both charged and 
neutral particles, and hence provides better event-plane resolution. The 
event-plane angle is determined by the event flow vector 
$Q_2$~\cite{MethodsQ}. The $Q$-vectors are recentered according to the 
procedure described in~\cite{MethodsQ}. The raw event-plane angle is 
estimated by:

\begin{equation}\label{raw_event_plane_angle}
 n \Psi^{\rm Raw}_{n} = \arctan{\frac{Q_{n,x}}{Q_{n,y}}},
\end{equation}

\noindent where $Q_{n,x}$ and $Q_{n,y}$ are the $x$ and $y$ projections 
of the flow vector. The flattening procedure described 
in~\cite{MethodsQ,Flattening} is applied to the $\Psi^{\rm Raw}_2$ 
distributions to remove detector acceptance effects. The resolution 
${\rm Res}(\Psi_2)$ values are evaluated using the three-subevent 
method~\cite{MethodsQ} correlating independent measurements made in 
the FVTX or MPC, BBC and the central arms (CNT) and are presented in 
Table~\ref{table:Res}.

\begin{table}[tbh]
\caption{\label{table:Res}
Values of the second-order event-plane resolution $\rm Res(\Psi_2)$ in \cuau
collisions at \sqsn = 200 GeV and \uu collisions at \sqsn = 193 GeV.}
\begin{ruledtabular}
\begin{tabular}{ccc}
Collisions & Centrality    & $\rm Res(\Psi_2)$ \\
\hline
\cuau      & 0\%--20\%     &      0.374        \\
           & 20\%--40\%    &      0.404        \\
           & 40\%--60\%    &      0.304        \\
           & 20\%--60\%    &      0.357        \\
\uu        & 0\%--50\%     &      0.495        \\
\end{tabular} 
\end{ruledtabular}
\end{table}

\subsection{The \vphi-meson raw yield extraction}

The yields of \vphi mesons ($N_{\vphi}$) are extracted by invariant-mass 
analysis via decay into oppositely charged kaons ($\vphi \rightarrow 
K^{+}K^{-}$). For $\vphi \rightarrow K^{+}K^{-}$ decay, the Particle 
Data Group~\cite{RevPartPhys} values are 
\begin{itemize}
\setlength\itemsep{0.1em}
\item mass = $1019.455\pm0.020$~\Mevcc,
\item width ($\Gamma$) = $4.26\pm0.04$~MeV, and
\item branching ratio = $48.9\pm0.5$~\%.
\end{itemize}

\noindent The analysis method follows a consolidated 
technique described extensively in 
Refs.~\cite{phi_AuAu,PhidAuCuCuAuAu,Flow_phi,Flow_pi0,Flow_pi0_eta}.

The measurements use two PHENIX central arms, each covering 
$|\eta|<0.35$ in pseudorapidity and $90^o$ in azimuthal angle. The 
central arms include a tracking system~\cite{TrackingSystem}, which 
comprises drift chambers and pad chambers. The tracking system is used 
for three-momentum-components determination for every track with a 
typical resolution of $\delta p/p = 0.7\%\oplus1.1\%\times p$ [GeV/$c$]. 
The time of flight ($\tau_f$) for hadrons is measured using the east-arm 
time-of-flight detector (TOFE)~\cite{TOF,TOF2} and the BBC. Information 
from the tracking system and $\tau_f$ allows for clear $\pi/K$ 
separation for $0.3<p_T<2.2$~\Gevc~\cite{phi_AuAu}.

\begin{figure}[ht]
    \includegraphics[width=0.96\linewidth]{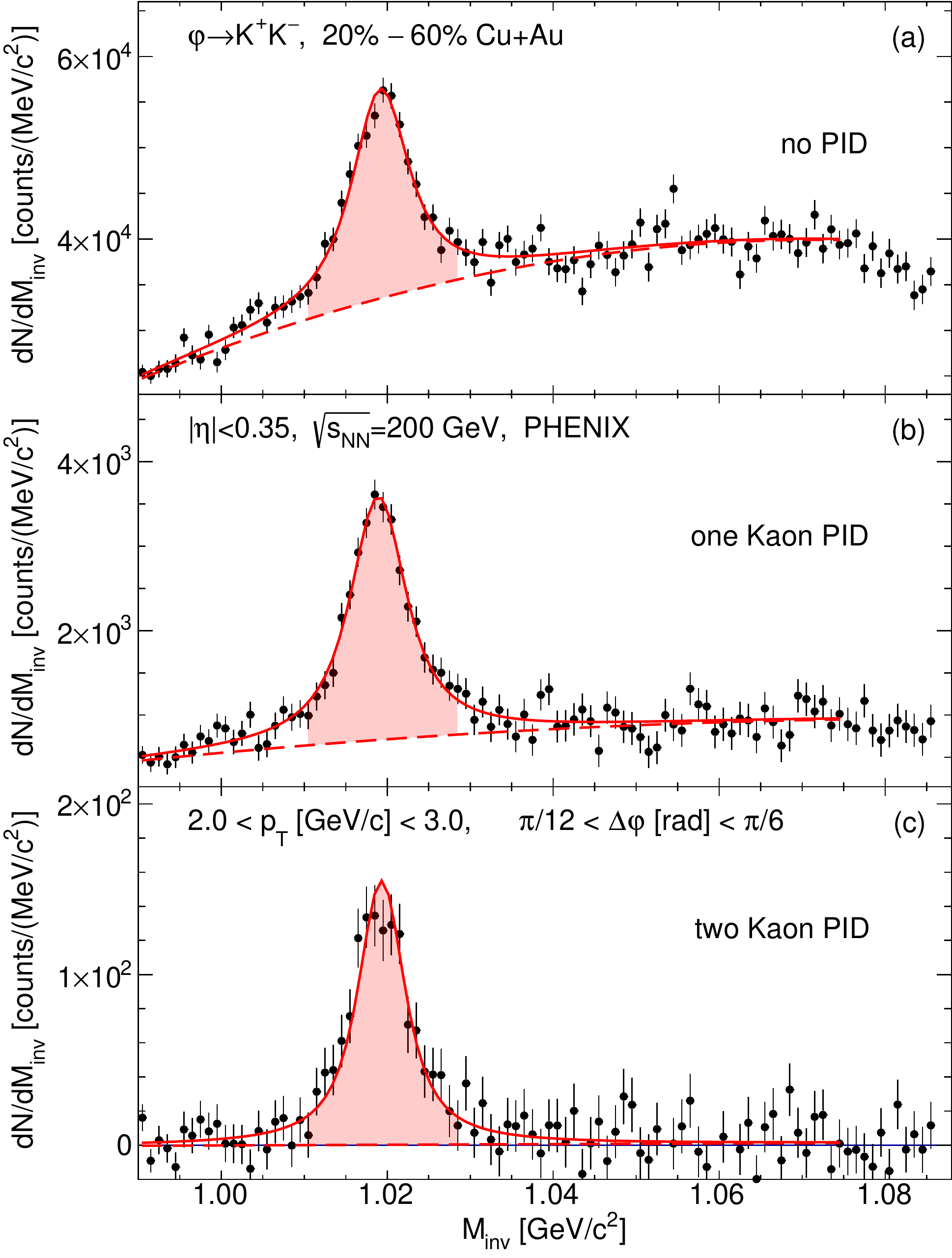}
\caption{
Shown are three examples of invariant-mass distributions for the 
$K^+K^-$ pairs in 20\%--60\% \cuau collisions at \sqsn = 200 GeV at 
$2.0<\pt<3.0$~\Gevc and $\pi/12<\Delta\vphi<\pi/6$~rad.  The 
distributions contain (a) no PID, (b) one-kaon PID, and (c) two-kaons 
PID methods after subtraction of the uncorrelated background estimated 
using the event-mixing technique.  Spectra are fitted to the sum of a 
Voigt function and polynomial of the third order, which described the 
\vphi-meson signal and the residual background, respectively.}
\label{fig:Minv}
    \end{figure}

In each event, all tracks of opposite charge that pass the selection 
criteria~\cite{phi_AuAu, PhidAuCuCuAuAu} are paired to form the 
invariant-mass distribution ($m_{KK}$) in the selected \vphi-meson \pt 
and event-centrality ranges. To maximize the statistical significance 
and the \pt reach of the measurements, three different pair-combination 
techniques are used. The first (``no PID'') does not require 
identification of charged tracks in the final state and assumes that all 
tracks are kaons. The second (``one-kaon PID'') requires identification 
of only one kaon in the TOFE subsystem. The third technique 
(``two-kaons PID'') identifies both kaons in the TOFE.  Each 
technique has advantages and disadvantages described in 
Ref.~\cite{PhidAuCuCuAuAu}. Both approaches with kaon identification 
have a more favorable signal-to-background ratio compared to the``no 
PID'' technique, but due to the small acceptance of the TOFE detector 
and its limited capability to identify kaons at $\pt>2.2$~\Gevc, these 
techniques have a limited \pt reach. The ``no PID'' approach allows the 
measurements to be extended towards higher \pt as it has substantially 
larger acceptance and a phase-space volume available for daughter kaons. 
All mentioned analysis techniques have a significant overlap in \pt and 
different sources of systematic uncertainties providing a valuable 
consistency check. The results obtained with different methods can not 
be directly averaged~\cite{PhidAuCuCuAuAu}. Therefore, to 
obtain the smallest statistical uncertainties for measurements of 
\vphi-meson \pt spectra and \vtwo values, the ``no PID'' approach is used at 
$\pt>2.2$ and $2.0$ \Gevc respectively, the ``one-kaon PID'' is used at lower 
\pt values, and the ``two-kaon PID'' is used for cross check and to estimate 
systematic uncertainties. Figure~\ref{fig:Minv} shows a typical 
invariant-mass distribution obtained using each of the three PID methods.

A large background that comes from random combinations of uncorrelated 
hadrons affects the invariant-mass spectrum. To estimate this background, 
a mixed-event technique~\cite{phi_AuAu} is applied that uses 
unlike-sign kaon tracks taken from different events with similar 
characteristics (i.e.~centrality and $z$-vertex). After subtraction, 
distributions are fitted with the sum of a Breit-Wigner mass-distribution 
function and a polynomial of the third order, which describe the 
\vphi-meson signal and the residual background, respectively.  The 
\vphi-meson raw yields are obtained as the integral of the mass 
distribution in a window of $\pm$9 MeV/c$^2$ ($\pm 2\Gamma$ 
\cite{Phi_raw_yields}) around the \vphi meson mass after subtracting the 
residual background.

\subsection{Invariant spectra and nuclear-modification factors}

The \pt-differential yields are corrected for the \vphi-meson 
reconstruction efficiency and acceptance of the detector, as described 
in~\cite{phi_AuAu,PhidAuCuCuAuAu}, using {\sc geant}3~\cite{GEANT} 
Monte-Carlo simulations for the 2012 configuration of the PHENIX 
detector.  The selection criteria for kaons and \vphi-meson candidates 
are the same in Monte Carlo and real data. The acceptance and 
reconstruction efficiency $\varepsilon_{\rm eff}(\pt)$ are evaluated as a 
ratio of reconstructed to generated \vphi mesons for the appropriate 
kinematic bin and event centrality in simulation.

Invariant transverse-momentum spectra of \vphi mesons are calculated as:
\begin{equation}\label{Spectra}
\frac{1}{2\pi\pt} \frac{d^2N}{d\pt dy}=\frac{1}{2\pi\pt} \frac{1}{N_{\rm event} Br} \frac{1}{\varepsilon _{\rm eff}(\pt)}\frac{N_{\vphi}(\Delta \pt)}{\Delta \pt \Delta y}
\end{equation}

\noindent where \pt is the transverse momentum, $\Delta$\pt is the 
transverse momentum interval, $\Delta y$ is the rapidity interval, and 
$N_{\rm event}$ is the number of events in the selected centrality bin.
Nuclear-modification factors (\rab) are used to study modifications to 
particle spectra~\cite{Mitrankova_2021} and are calculated as

\begin{equation}\label{RxA}
R_{AB}=\frac{\sigma^{\rm inel}_{pp}}{\langle N_{\rm coll} \rangle}\cdot \frac{d^2N_{AB}/dydp_T}{d^2\sigma_{pp}/dydp_T},
\end{equation}

\noindent where $d^2N_{AB}/dydp_T$ is the per-event yield of particle 
production in $A$$+$$B$ collisions, $d^2\sigma_{pp}/dydp_T$ is the 
production cross section in \pp collisions, and 
$\sigma^{\rm inel}_{pp}=42.2$ mb~\cite{pp} is the total inelastic cross 
section in \pp collisions.

\subsection{Elliptic Flow}

A robust method~\cite{Flow_phi,Flow_pi0_eta,Flow_pi0} is used to study 
the elliptic flow of resonance particles, such as the \vphi meson. To 
obtain the azimuthal-angle dependence of \vphi-meson production, the 
\vphi-meson raw yields are measured in a selected \pt range as a 
function of the $K^{+}K^{-}$ pair angle with respect to the 
reaction-plane orientation in six equally spaced bins of 
$\Delta\varphi=\varphi_{\rm pair}-\Psi_{2}$ covering the range 
$0<\Delta\varphi<\pi/2$. Assuming elliptic flow is the dominant source of 
the $\Delta\varphi$ variation in the \vphi-meson 
yields~\cite{Flow_phi,Flow_pi0,Flow_pi0_eta}, the \vtwo coefficients are 
then extracted from a fit to the distribution 
$dN_{\vphi}/d(\Delta\varphi)$ using the function~\cite{Flow_meth}:

\begin{equation}\label{fit_for_v2}
\frac{dN_{\vphi}}{d(\Delta\varphi)}=M(1+2\vtwo^{\rm obs}\cos{[2\Delta\varphi]}),
\end{equation}

\noindent where $M$ is a normalization constant. Because of the finite 
bin width in $\Delta\varphi$, the extracted $\vtwo^{\rm obs}$ values are 
corrected by a smearing factor $\sigma = \delta/\sin{\delta}$, which 
accounts for the finite bin width $\delta=\pi/12$. The \vtwo extractions 
are performed for all of the aforementioned PID approaches and the 
results with the smallest statistical uncertainties are used in the 
analysis.  The final \vphi-meson \vtwo values are evaluated as 

$\vtwo=\vtwo^{\rm obs}/{\rm Res}(\Psi_2)$.

An alternative method to evaluate \vphi-meson \vtwo is the invariant 
mass fit method, described 
in~\cite{Flow_phi,Flow_meth,MethodsQ,Flow_pi0_eta}. In this 
analysis it is used to perform cross check and for the evaluation of 
systematic uncertainties.


\begin{table*}[tb]
\caption{\label{table:syst_v2}
Values of systematic uncertainties (\%) for \vphi-meson \vtwo 
measured in \cuau collisions at \sqsn = 200 GeV and \uu 
collisions at \sqsn = 193 GeV.}
\begin{ruledtabular}
\begin{tabular}{cccccc}
                 &                 \multicolumn{4}{c}{\cuau}              &  \uu    \\
Uncertainty      & 0\%--20\%   & 20\%--40\% & 40\%--60\% & 20\%--60\% & 0\%--50\%   \\
\hline
Reaction plane   & 7.0         & 2.0        & 3.0        & 1.0        & 3.0         \\
Acceptance       & 3.0         & 3.0        & 3.0        & 3.0        & 3.0         \\
Yield extraction & 9.8--12.5  & 11.4--17.3  & 12.2--14.1 & 8.4--14.0  & 10.9--13.2  \\
Total            & 12.8--14.9 & 12.3--17.9  & 13.3--15.1 & 9.4--14.7  & 12.1--14.2  \\
\end{tabular}
\end{ruledtabular}
\end{table*}

\begin{table}[tbh]
\caption{\label{table:Syst_cuau}
Summary of systematic uncertainties (\%) on the \vphi-meson
invariant yields in \cuau collisions at \sqsn = 200 GeV.}
\begin{ruledtabular}
\begin{tabular}{cccc}
                                & \multicolumn{3}{c}{\pt (\Gevc)}  \\
Uncertainty                     & 1.45  &3.45  & 7.00       \\
\hline
Acceptance                      & 4.5   & 3.0    & 3.0  \\
Peak extraction 0\%--93\%       & 6.2   & 8.5  & 12.6  \\
Peak extraction 0\%--20\%       & 7.6   & 10.7 & 16.2  \\
Peak extraction 20\%--40\%      & 7.9   & 10.1 & 14.1  \\
Peak extraction 40\%--60\%      & 9.6   & 11.2 & 14.0  \\
Peak extraction 60\%--80\%      & 8.3   & 12.5 & 19.9  \\
Reconstruction efficiency       &3.0    & 3.0  &  3.0  \\
Momentum scale                  & 0.6   & 3.0  & 5.0   \\
Branching ratio                 & 1.2   & 1.2  & 1.2   \\
Total 0\%--93\%                 & 7.9   & 10.2 & 14.4  \\
Total 0\%--20\%                 & 9.0   & 12.1 & 17.6  \\
Total 20\%--40\%                & 9.3   & 11.6 & 15.7  \\
Total 40\%--60\%                & 10.8  & 12.6 & 15.6  \\
Total 60\%--80\%                & 9.6   & 13.7 & 21.1  \\
\end{tabular}
\end{ruledtabular}
\end{table}

\begin{table}[tbh]
\caption{\label{table:Syst_uu}
Summary of systematic uncertainties (\%) on the \vphi-meson
invariant yields in \uu collisions at \sqsn = 193 GeV}
\begin{ruledtabular}
\begin{tabular}{cccc}
                            &  \multicolumn{3}{c}{\pt (\Gevc)} \\
Uncertainty                 & 1.10 & 3.45 & 7.00      \\
\hline
Acceptance                  & 4.0  & 3.0  & 3.0  \\
Peak extraction 0\%--20\%   & 10.8 & 7.2  & 18.5 \\
Peak extraction 20\%--40\%  & 11.3 & 8.1  & 16.9 \\
Peak extraction 40\%--60\%  & 13.6 & 7.1  & 15.8 \\
Peak extraction 60\%--80\%  & -    & 9.7  & 20.5 \\
Reconstruction efficiency   & 2.5  & 2.0  & 2.0  \\
Momentum scale              & 0.5  & 3.6  & 5.0  \\
Branching ratio             & 1.2  & 1.2  & 1.2  \\
Total 0\%--20\%             & 11.8 & 8.6  & 19.7 \\
Total 20\%--40\%            & 12.2 & 9.4  & 18.1 \\
Total 40\%--60\%            & 14.4 & 8.5  & 16.0 \\
Total 60\%--80\%            & -    & 10.8 & 21.0  \\
\end{tabular}
\end{ruledtabular}
\end{table}

\subsection{Systematic uncertainties}

The calculation of the systematic uncertainties follows the procedure 
performed in~\cite{PhidAuCuCuAuAu,Flow_phi,phi_AuAu,FlowCuAu}. The main 
sources of systematic uncertainties for \vphi-meson \vtwo and \pt 
spectra are summarized in 
Tables~\ref{table:syst_v2}-\ref{table:Syst_uu}.
Systematic uncertainties are grouped into three types:
\begin{itemize}
\setlength\itemsep{0.1em}
\item[A:] (point-to-point uncorrelated), which can move 
each point independently;
\item[B:] (point-to-point \pt-correlated), which can move 
points coherently, but not necessarily by the same relative amount;
\item[C:] (global), which move all points by the same 
relative amount.
\end{itemize}

The main contribution to the systematic uncertainties of type A is the 
uncertainty in the raw-yield extraction of 6\%-20\%, evaluated by 
varying the identification approaches, fit parameters and the 
parameterization of the residual background. An uncertainty of type B is 
dominated by uncertainties in acceptance of 3\%-4.5\%, reconstruction 
efficiency $\varepsilon_{\rm rec}$ of 2.0\%-3.0\%, and momentum scale 
0.5\%-5.0\%. The main contributions to the type C uncertainties are the 
uncertainties in normalization for the cross section (9.7\%) and \Ncoll 
calculations presented in Table \ref{table:Ncoll_Npart_Cuau_UU}.

The systematic uncertainties of type A for $\vtwo^{\rm obs}$ of 
8.4\%-14.0\% are estimated by varying the elliptic flow measurement 
method, identification cuts for \vphi mesons, the parameterization of 
the residual background, and the peak integration window in the $m_{KK}$ 
distributions. The \vphi-meson \vtwo systematic uncertainties of type B 
and C have two main sources: acceptance (3\%) and reaction plane 
determination (1.0\%-7.0\%).

For \vphi-meson \pt spectra and \rab, the systematic
uncertainties of type A and B are added in quadrature to
give the total systematic uncertainties.  For \vphi-meson
\vtwo, all uncertainties are added in quadrature to give
the total systematic uncertainties.

\section{RESULTS}

\begin{figure}[htb]
\includegraphics[width=1.\linewidth]{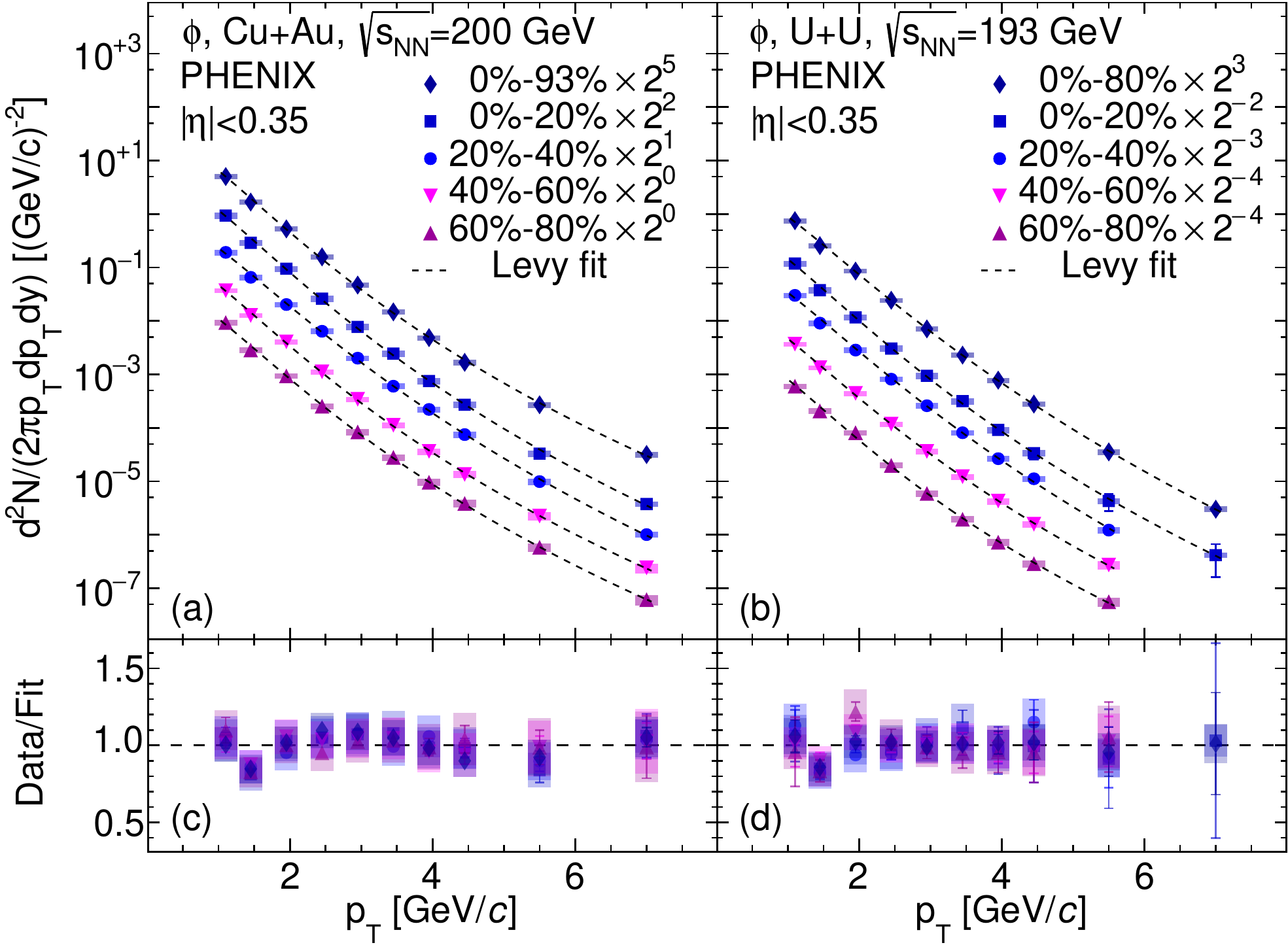}
\caption{The invariant transverse momentum spectra measured for \vphi 
mesons in (a) \cuau at \sqsntwo and (b) \uu collisions at \sqsnuu at 
midrapidity. The statistical uncertainties are represented by vertical 
lines (hidden by the markers) while the systematic uncertainties 
represented by rectangles around the points. Panels (c) and (d) show 
data-to-fit ratios.}
\label{fig:Spectra}
\end{figure}

\subsection{Invariant transverse-momentum spectra}

Figure~\ref{fig:Spectra} shows the invariant \pt spectra 
of \vphi mesons measured in (a) \cuau collisions at \sqsntwo and (b) \uu 
collisions at \sqsnuu at midrapidity \midrap. The \vphi-meson spectra 
are measured from 1.1 to 7.0 \Gevc in \pt for 5 centrality classes in 
\cuau and \uu collisions.

The dashed lines on panels (a) and (b) of Fig.~\ref{fig:Spectra} 
represent the Levy function fits~\cite{Levy}:
\begin{multline}
\frac{1}{2\pi\pt} \frac{d^2N}{d\pt dy}=\frac{1}{2\pi} \frac{dN}{dy} \frac{(n-1)(n-2)}{nT(nT+m_{\vphi}(n-2))} \times \\	\times \left(1+\frac{\sqrt{\pt^{2}+m_{\vphi}^{2}}-m_{\vphi}}{nT}\right)^{-n},
\end{multline}

\noindent where $m_{\vphi}$ is the \vphi-meson mass, and $dN/dy$, $T$, 
and $n$ are free parameters. The $dN/dy$ term corresponds to the 
\vphi-meson multiplicity at midrapidity. The Levy function includes both 
an exponential shape for low \pt (which can be characterized by an 
inverse-slope parameter $T$) and a power-law component (governed by the 
power parameter $n$) for the higher \pt region. Panels (c) and (d) of 
Fig.~\ref{fig:Spectra} show data-to-fit ratios with fit function values 
taken at the bin center, and indicate good agreement between the 
measured \vphi-meson \pt spectra and the Levy function.


\subsection{Nuclear-modification factors}

\begin{figure}[ht]
\begin{minipage}[b]{1\linewidth}
\centering
\includegraphics[width=0.96\linewidth]{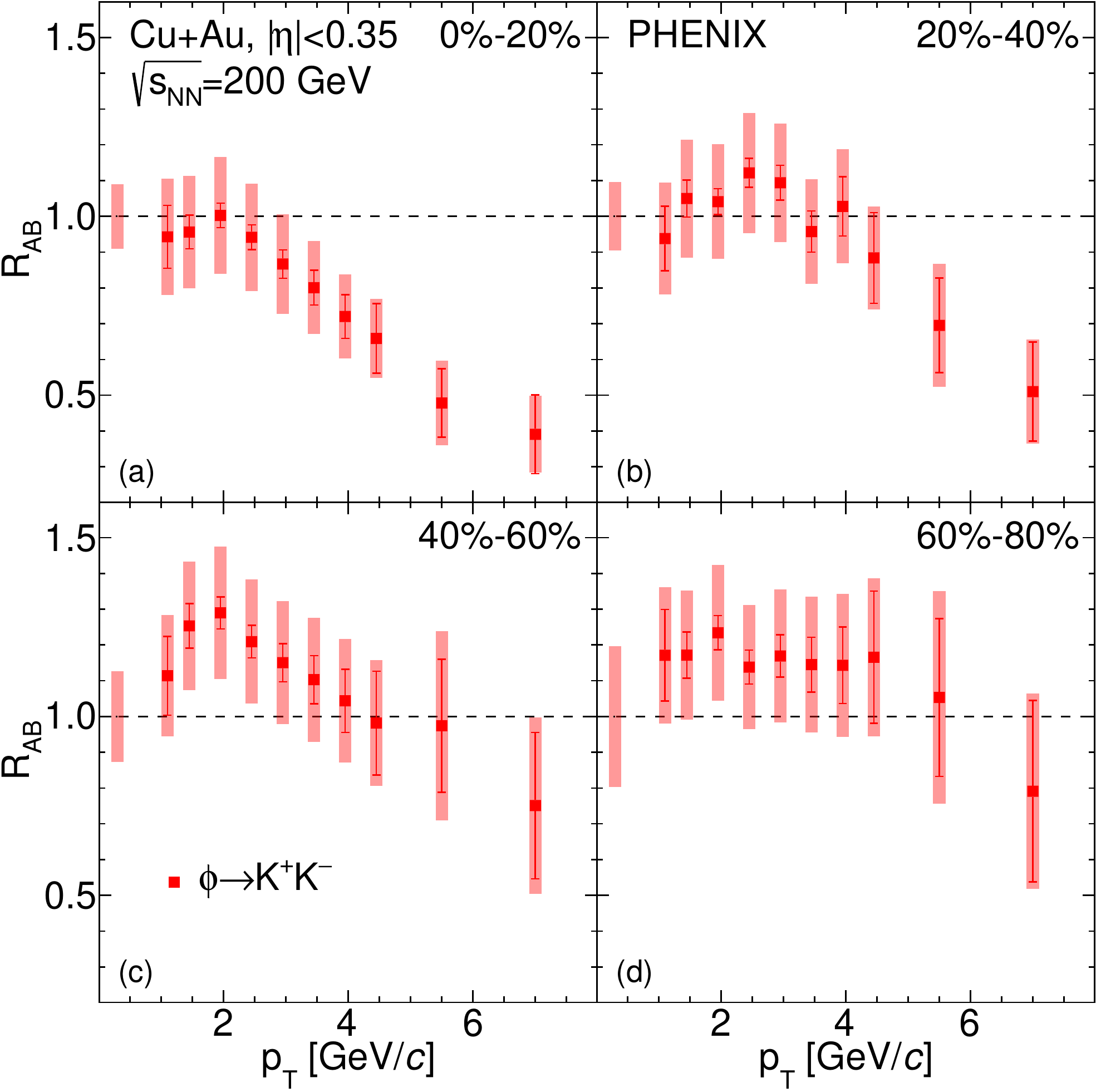}
\end{minipage}
\begin{minipage}[b]{1\linewidth}
\centering
\includegraphics[width=0.96\linewidth]{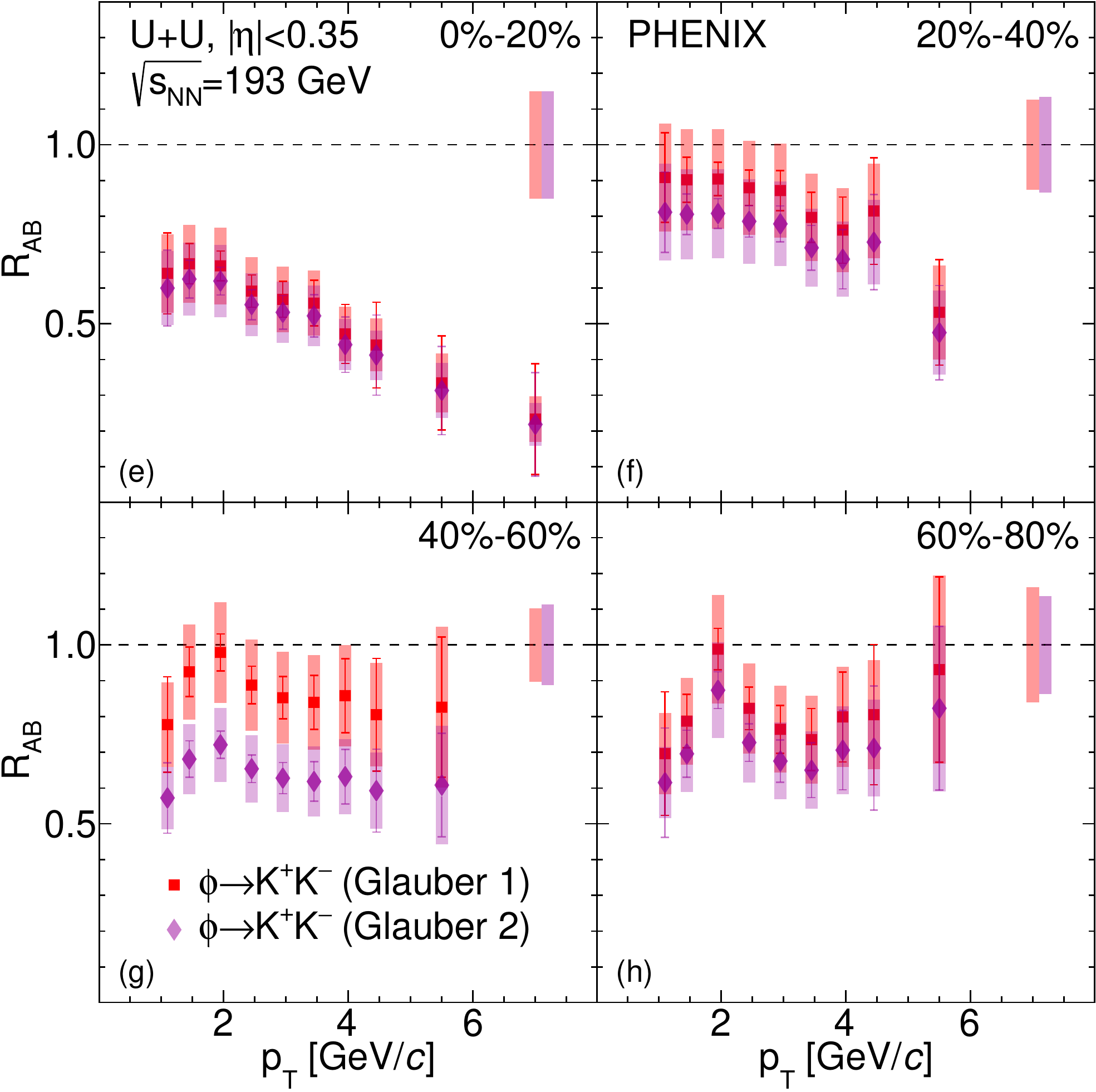}
\end{minipage}
\caption{The \vphi-meson nuclear-modification factors \rab measured as a 
function of \pt in different centrality intervals of (a) to (d) \cuau 
collisions at \sqsntwo and (e) to (h) \uu collisions at \sqsnuu at 
midrapidity $|\eta|<0.35$. The normalization uncertainty from \pp of 
about $9.7\%$ is not shown. Here and below the type C uncertainties are 
shown as boxes near unity.}
\label{fig:RAB}
\end{figure}

Figure~\ref{fig:RAB} shows \vphi-meson \rab measured in \cuau collisions 
at \sqsntwo and \uu collisions at \sqsnuu at midrapidity $|\eta|<0.35$. 
The reference \vphi-meson production cross section in \pp collisions is 
taken from~\cite{pp}. The normalization uncertainty from \pp is not 
shown. The \vphi-meson \rab values in central and semicentral \cuau and 
\uu collisions at high \pt$>5$ \Gevc are less than unity, indicating 
suppression. The high-\pt suppression of \vphi-meson yields decrease
when moving to more peripheral collisions.  The similar behavior of \vphi 
meson production has been observed in symmetric systems and is 
interpreted as indicative of in-medium jet 
quenching~\cite{phi_AuAu,PhidAuCuCuAuAu}. In the most peripheral \cuau 
and \uu collisions in whole \pt range, \vphi-meson \rab
factors values are close to unity within uncertainties.

\begin{figure}[ht]
\includegraphics[width=0.96\linewidth]{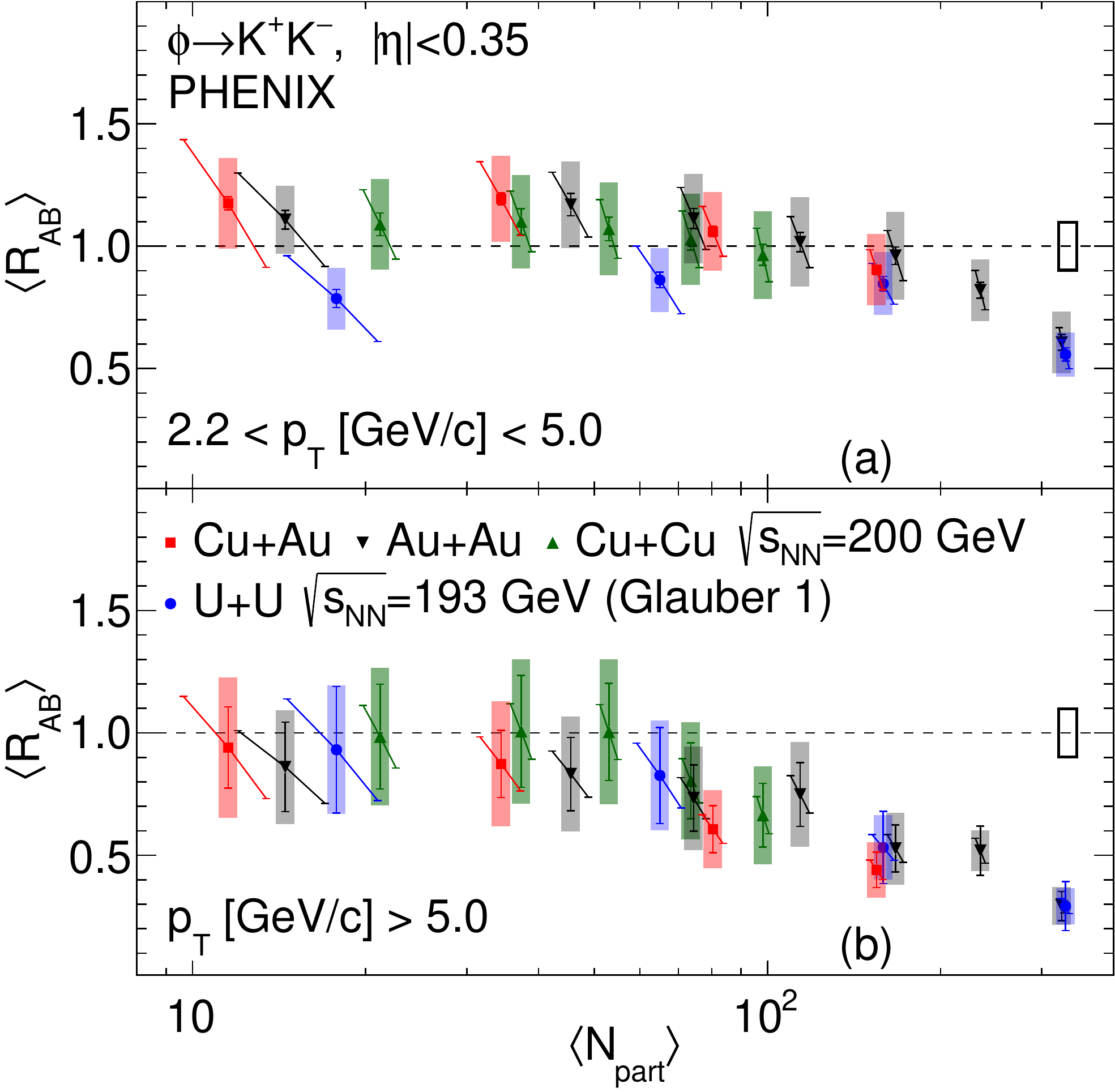}
\caption{The \vphi-meson integrated nuclear-modification factors 
$\langle{\rab}\rangle$ measured as a function of \Npart in 
\cucu~\cite{PhidAuCuCuAuAu}, \cuau, \auau~\cite{PhidAuCuCuAuAu} 
collisions at \sqsntwo, and \uu collisions at \sqsnuu integrated in (a) 
$2.2 < \pt < 5.0$ \Gevc and (b) $\pt > 5.0$ \Gevc at midrapidity 
$|\eta|<0.35$. The tilted bars represent correlated uncertainties from 
Glauber-Monte-Carlo simulation.}
\label{fig:RAB_INT}
\end{figure}

To better understand the features of \vphi-meson production, the 
integrated nuclear-modification factors $\langle{\rab}\rangle$ for \vphi 
mesons as a function of \Npart are shown in Fig.~\ref{fig:RAB_INT} for 
different collision systems (\cuau, \auau, and \cucu collisions at 
\sqsntwo, and \uu collisions at \sqsnuu). The \auau and \cucu results 
are taken from~\cite{PhidAuCuCuAuAu}.  The integrated 
$\langle{\rab}\rangle$ values were calculated as the averaged \rab 
values in the intermediate-\pt range ($2.2<\pt<5.0$~GeV/$c$) and in the 
high-\pt range ($\pt>5.0$~GeV/$c$), according to the procedure 
previously used in Refs.~\cite{Pi_eta_cuau,Pi_eta_uu,PhidAuCuCuAuAu}. 
The $\langle{\rab}\rangle$ values for \vphi mesons vs~\Npart obtained in 
the large collision systems are consistent within uncertainties, as has 
already been observed for \pio and \et mesons~\cite{Pi_eta_cuau, 
Pi_eta_uu}. The value of \Npart characterizes the volume of the 
nuclear-overlap area and hence is assumed to be proportional to the 
volume of the hot and dense matter formed in heavy ion 
collisions~\cite{FlowCuCuAuAu}. For that reason, the obtained 
$\langle{\rab}\rangle$ results suggest the scaling of light-hadron 
production integrated over azimuthal angle with the average 
nuclear-overlap size, regardless of the collision geometry.

\begin{figure*}[htp]
\begin{minipage}{0.99\linewidth}
\includegraphics[width=0.49\linewidth]{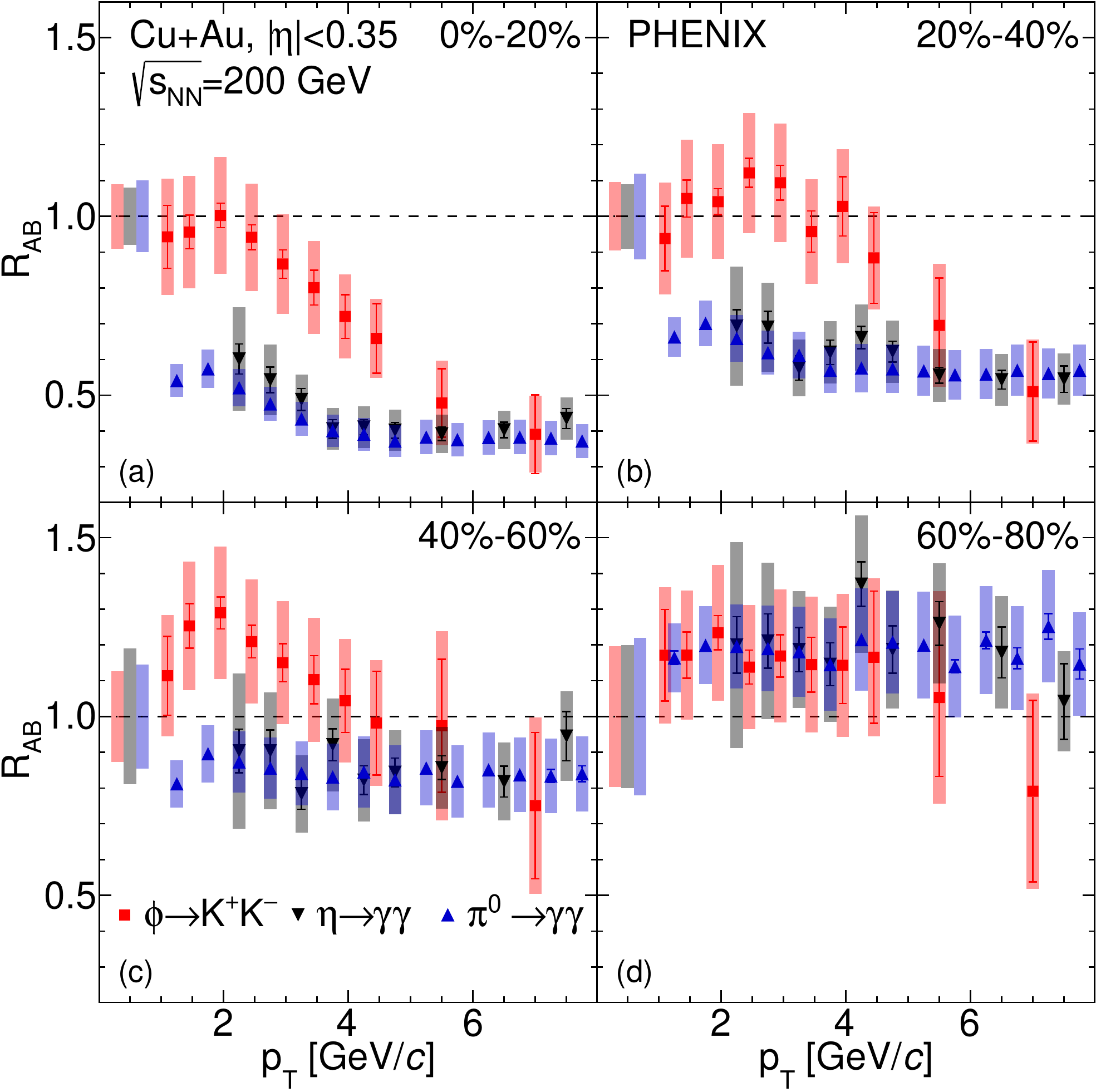}
\includegraphics[width=0.49\linewidth]{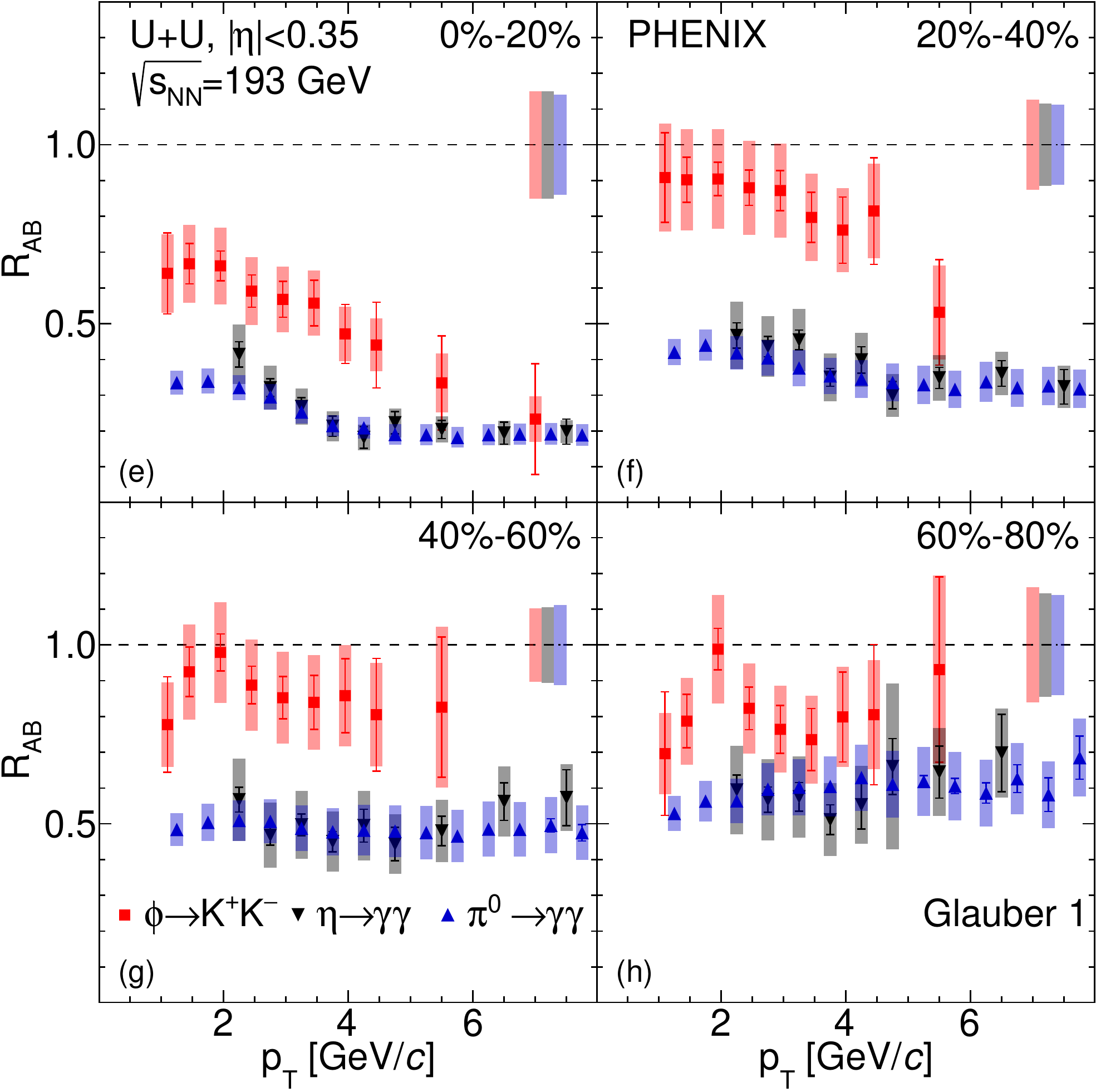}
\vspace{-0.2cm}
\caption{The comparison of \vphi-meson \rab values to \pio and \et meson 
\rab values measured as a function of \pt in different centrality 
intervals of (a) to (d) \cuau collisions at \sqsntwo and (e) to (h) \uu 
collisions at \sqsnuu at midrapidity. The \rab values for \pio and \et 
meson \rab in \cuau and \uu collisions are from~\cite{Pi_eta_cuau, 
Pi_eta_uu}.}
\label{fig:RAB_pi_eta}
\vspace{0.3cm}
\end{minipage}
\begin{minipage}{0.99\linewidth}
\includegraphics[width=1\linewidth]{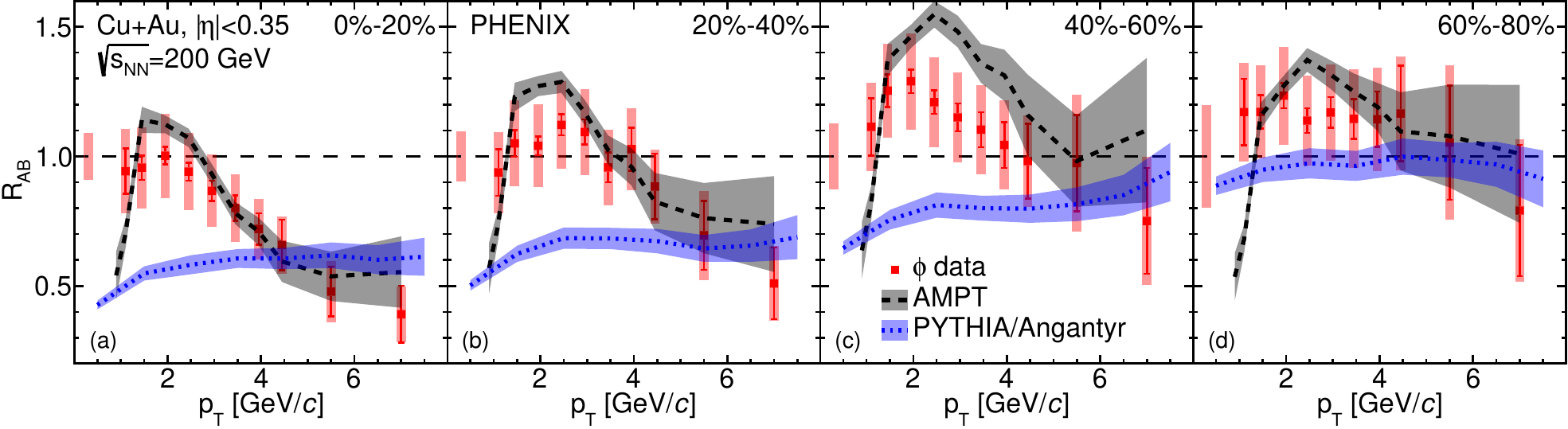}
\vspace{-0.2cm}
\caption{The comparison of \vphi-meson \rab values measured as a 
function of \pt in different centrality intervals of \cuau collisions at 
\sqsntwo at midrapidity ($|\et|<0.35$) to AMPT model and PYTHIA/Angantyr 
model predictions.}
\label{fig:AMPT_PYTHIA}
\vspace{0.3cm}
\end{minipage}
\end{figure*}

\begin{figure*}[htp]
\includegraphics[width=0.99\linewidth]{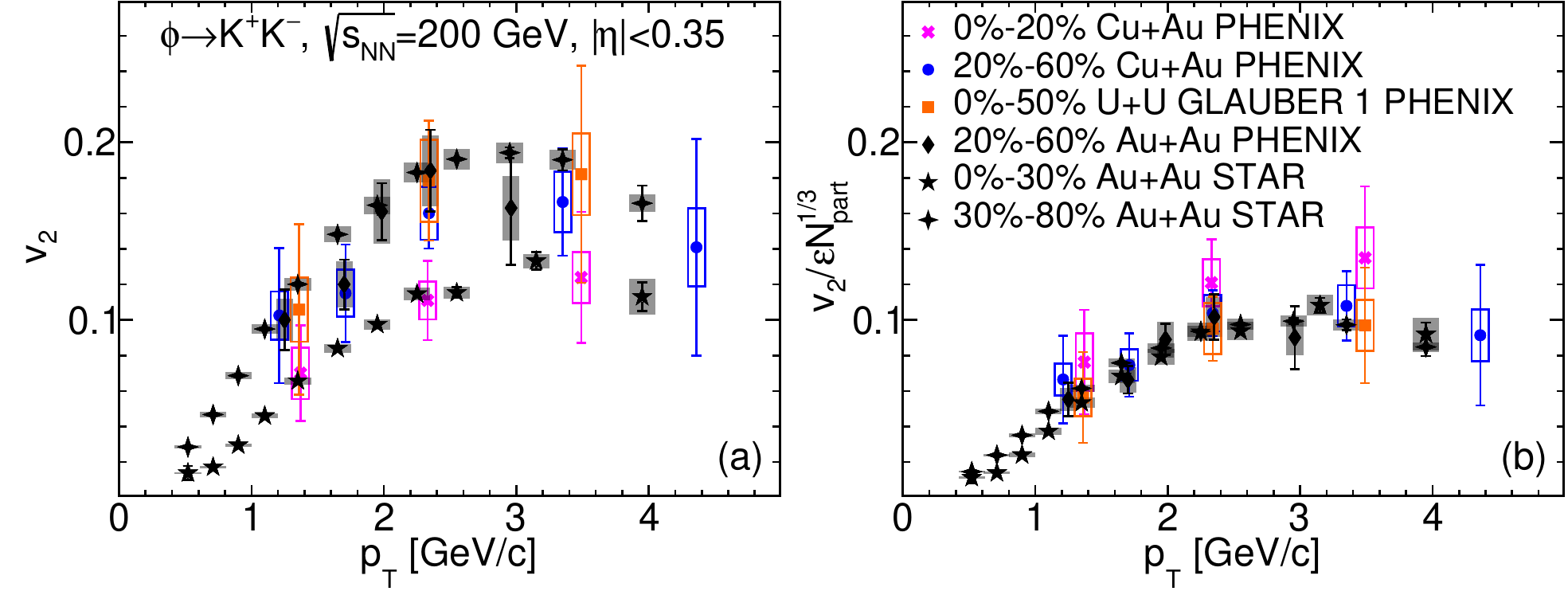}
\caption{The comparison of $\vphi$ meson (a) $\vtwo$ and (b) 
$\vtwo/(\varepsilon N^{1/3}_{\rm part})$ measured as a function of \pt 
in $\cuau$, $\uu$ and $\auau$~\cite{Flow_phi, STAR_v2_phi_AuAu} 
collisions at $\sqsntwo$ at midrapidity ($|\et|<0.35$). }
\label{fig:Phi_v2_all_syst}
\end{figure*}

\begin{figure*}[htb]
\begin{minipage}{0.99\linewidth}
\includegraphics[width=0.99\linewidth]{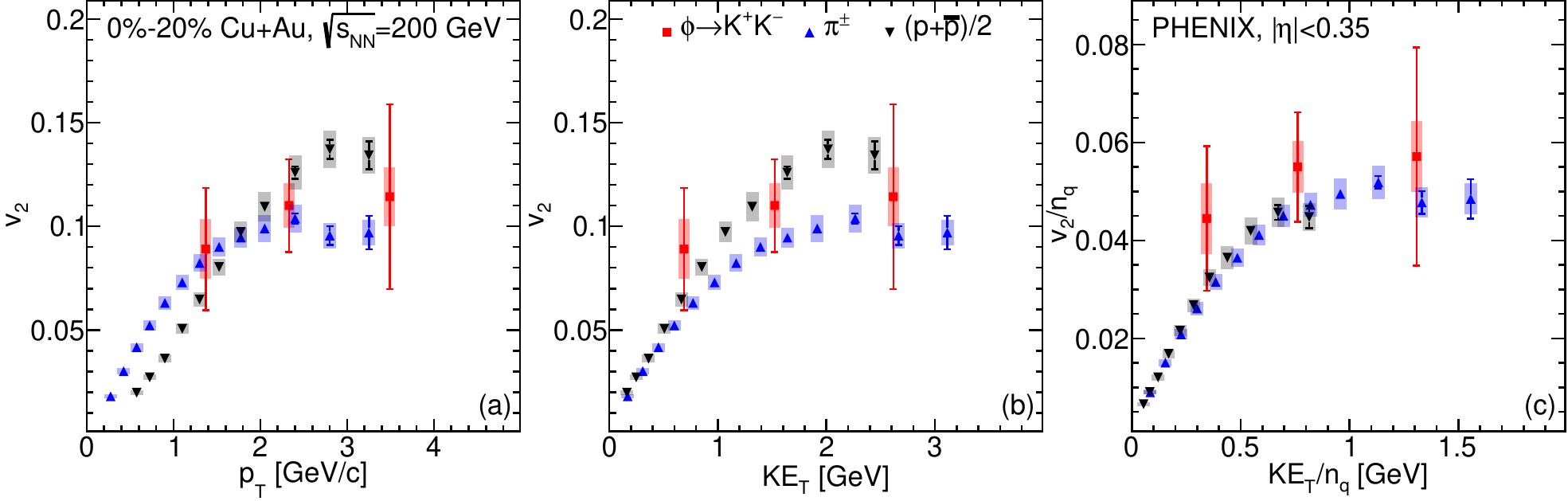}
\caption{\label{fig:Flow_CuAu_phi_pi0_prot_centr} The comparison of 
elliptic flow (a, b) \vtwo and (c) $\vtwo/\nq$ values measured for \vphi 
mesons as a function of (a) \pt, (b) \kEt and (c) $\kEt/\nq$ in 
0\%--20\% \cuau collisions to corresponding \vtwo and $\vtwo/\nq$ values 
for $\pi^\pm$ mesons and (anti)protons ($(p+\bar{p})/2$). The values for 
$\pi^\pm$ mesons and $(p+\bar{p})/2$ \vtwo are taken 
from~\cite{FlowCuAu}. }
\end{minipage}
\begin{minipage}{0.99\linewidth}
\vspace{0.5cm}
\includegraphics[width=0.99\linewidth]{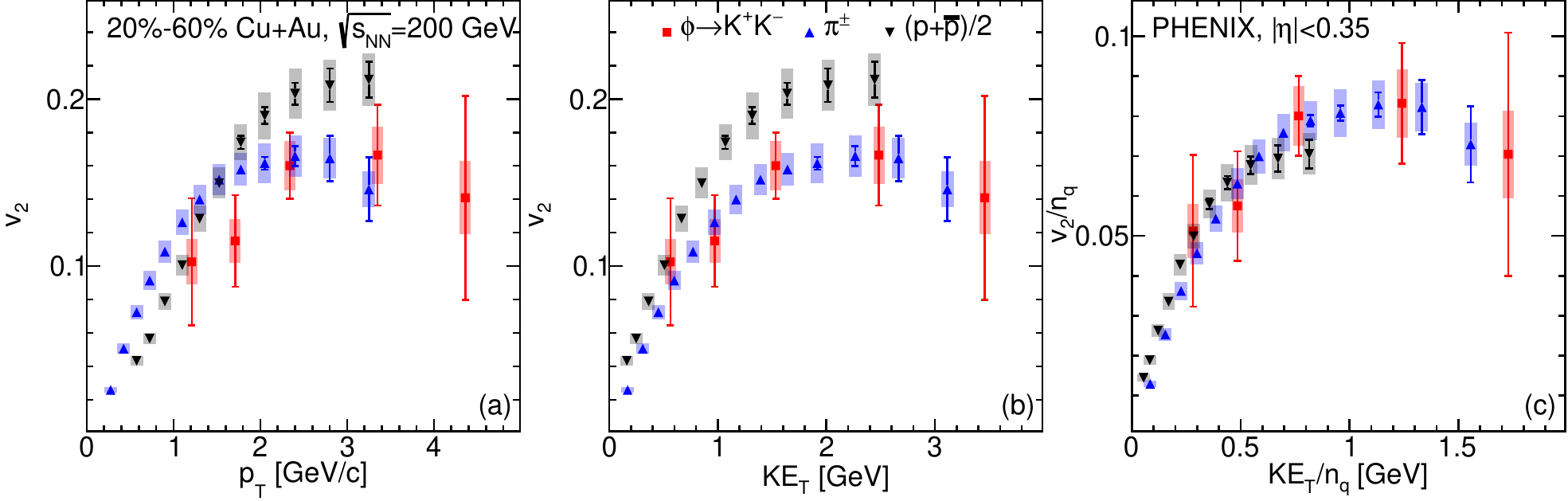}
\caption{\label{fig:Flow_CuAu_phi_pi0_prot_periph} The comparison of 
elliptic flow (a, b) \vtwo and (c) $\vtwo/\nq$ for \vphi mesons measured 
as a function of (a) \pt, (b) \kEt and (c) $\kEt/\nq$ in 20\%--60\% 
\cuau collisions to corresponding \vtwo and $\vtwo/\nq$ values for 
$\pi^\pm$ mesons and (anti)protons ($(p+\bar{p})/2$). The values for 
$\pi^\pm$ mesons and $(p+\bar{p})/2$ \vtwo are taken 
from~\cite{FlowCuAu}.}
\end{minipage}
\begin{minipage}{0.99\linewidth}
\vspace{0.5cm}
\includegraphics[width=0.99\linewidth]{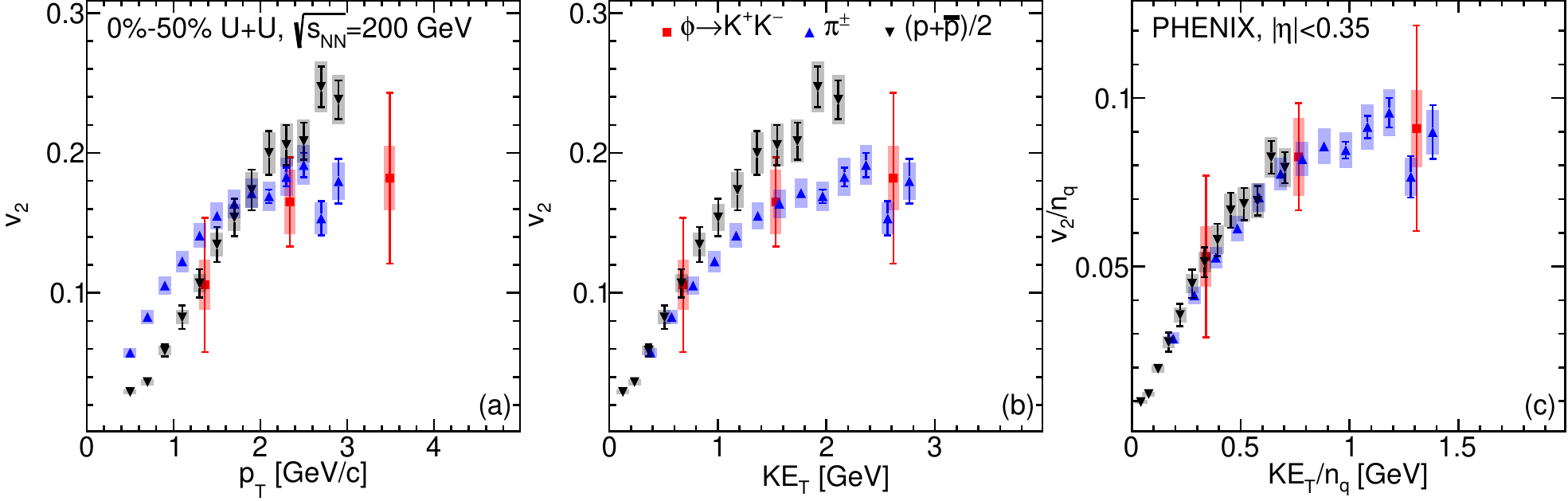}
\caption{\label{fig:Flow_UU_phi_pi0_prot} The comparison of elliptic 
flow (a, b) \vtwo and (c) $\vtwo/\nq$ for \vphi mesons measured as a 
function of (a) \pt, (b) \kEt and (c) $\kEt/\nq$ in 0\%--50\% \uu 
collisions to corresponding \vtwo and $\vtwo/\nq$ values for $\pi^\pm$ 
mesons and (anti)protons ($(p+\bar{p})/2$).The values for $\pi^\pm$ 
mesons and $(p+\bar{p})/2$ \vtwo are taken from~\cite{v2_UU}.}
\end{minipage}
\end{figure*}

\begin{figure*}[htp]
\includegraphics[width=0.99\linewidth]{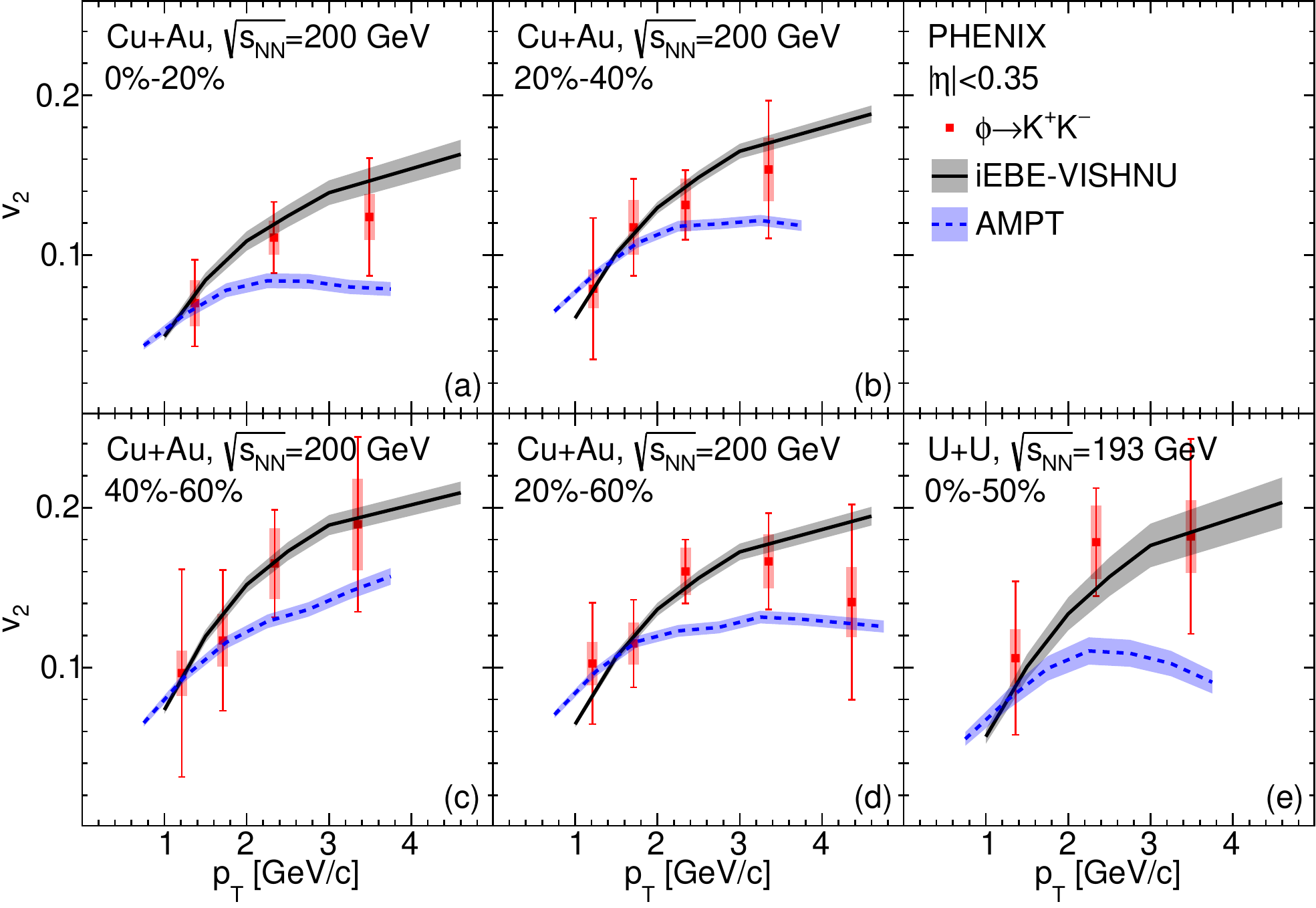}
\caption{The comparison of elliptic flow \vtwo(\pt) for \vphi mesons 
measured in
(a) 0\%--20\%,
(b) 20\%--40\%,
(c) 40\%--60\%, and
(d) 20\%--60\% \cuau collisions and
(e) 0\%--50\% \uu collisions to iEBE-VISHNU hydrodynamic model 
predictions with specific viscosity $\et/s=1/(4\pi)$ and AMPT model 
predictions.}
\label{fig:Flow_iEBE}
\end{figure*}


Figure~\ref{fig:RAB_pi_eta} shows the comparisons of \vphi-meson \rab 
values to \pio and \et meson \rab values~\cite{Pi_eta_cuau, Pi_eta_uu} 
obtained in \cuau collisions at \sqsntwo and \uu collisions at \sqsnuu 
at midrapidity.  The \vphi-meson \rab values are larger than \pio and \et 
meson \rab values in the central collisions in the intermediate-\pt 
range. The differences between \pio and \et meson \rab and \vphi-meson 
\rab values at moderate \pt decrease as the centrality increases. These 
trends of light-hadron production in the intermediate \pt range can be 
qualitatively explained in terms of the interplay of strangeness 
enhancement and hadronization via coalescence~\cite{Recombination_phi}. 
In central \cuau and \uu collisions at high \pt, all light meson yields 
show the same suppression level. The high \pt suppression is 
consistent---within the measurement uncertainties---with the assumption 
of flavor-independent energy loss of pre-fragmentation partons ($u,d,s$ 
quarks) in the hot and dense medium. The same light-hadron \rab behavior 
has been observed in symmetric \cucu and \auau 
collisions~\cite{PhidAuCuCuAuAu}, indicating that features of 
light-hadron production do not depend on collision geometry.

Figure~\ref{fig:AMPT_PYTHIA} presents comparisons of \vphi-meson \rab 
values, measured in \cuau collisions at \sqsntwo, to \rab values 
estimated with AMPT~\cite{AMPT} and 
PYTHIA/Angantyr~\cite{PYTHIA_ANGANTYR} models. The version of the AMPT 
model employed here includes the string-melting 
mechanism~\cite{StringMelting}. String melting refers to excited 
strings, i.e.~those not coming from projectile and target nucleons that 
do not interact, which are converted (``melted'') into partons. Those 
produced partons undergo some small number of scatterings and then 
coalesce (using a simple spatial-coalescence mechanism) into hadrons.

In contrast, the PYTHIA/Angantyr model comprises a coherent set of physics 
models for the evolution from a few-body hard-scattering process to a 
complex-multiparticle final state~\cite{PYTHIA8}. The \vphi-meson \rab 
values in both models are obtained by treating the model outputs in the 
same way as the experimental data. For \rab calculation with the AMPT 
model, the \vphi-meson production cross section measured in \pp 
collisions is used as a baseline. To calculate \vphi-meson \rab with 
PYTHIA/Angantyr model, the \vphi-meson \pt yields in \cuau collision 
obtained in PYTHIA/Angantyr are divided by \vphi-meson \pt yields in \pp 
collision from PYTHIA8, and by the same \Ncoll as in experiment. The 
AMPT results are obtained using a parton-scattering cross section of 3.0 
mb and incorporating the nuclear-shadowing effect~\cite{AMPT}. The 
parameters used in the event generation of PYTHIA/Angantyr are listed in 
Table~\ref{table:PYTHIA_parameters}. The multiplication factor for 
multiparton interactions is introduced to match charged hadron 
multiplicities in \pp collisions at \sqsn = 200~GeV in PYTHIA 
calculations and experimental data~\cite{dNch_deta}. PYTHIA/Angantyr 
calculations include uncertainties estimated from the variation of 
parton-distribution functions.

To quantify the agreement of model calculations with experimental 
results, $p$-values~\cite{p_val} are calculated from the least-squares 
minimization in the standard way. Table~\ref{table:p_val_RAB_CuAu} shows 
the $p$-values estimated for the string melting version of AMPT and 
PYTHIA/Angantyr model calculations of \vphi-meson \rab values in 
different centrality classes of \cuau collisions at \sqsn = 200~GeV at 
midrapidity.

The values of \vphi-meson \rab measured in peripheral \cuau collisions 
are well described by both PYTHIA/Angantyr and AMPT model calculations. 
In the most-central and semicentral \cuau collisions at moderate \pt, 
\vphi-meson \rab values obtained with PYTHIA/Angantyr are significantly 
smaller than the measured \rab values, whereas the AMPT model reproduces 
the measured \vphi-meson \rab values as reasonably well supported by the 
calculated $p$-values (Table~\ref{table:p_val_RAB_CuAu}). This means 
\vphi-meson production measured in \cuau collisions is well described by 
the AMPT model, which assumes that the mechanism of \vphi-meson production 
at moderate \pt is dominated by the coalescence of $s\bar{s}$ 
pairs~\cite{AMPT}.


\subsection{Elliptic flow}

Figure~\ref{fig:Phi_v2_all_syst} presents \vphi-meson \vtwo values as a 
function of \pt measured in 0\%--20\% and 20\%--60\% \cuau collisions 
and 0\%--50\% \uu collisions. Elliptic flow values for \vphi mesons 
previously obtained in 20\%--60\% \auau collisions by 
PHENIX~\cite{Flow_phi} and in 0\%--30\% and 30\%--80\% \auau collisions 
by STAR~\cite{STAR_v2_phi_AuAu} are also shown in 
Fig.~\ref{fig:Phi_v2_all_syst}. The comparison of elliptic flow for 
\vphi mesons in symmetric and asymmetric collision systems suggests that 
the \vtwo values follow common empirical scaling with 
$\varepsilon_2 N_{\rm part}^{1/3}$. Scaling with participant 
eccentricity of second-order $\varepsilon_2$ represents dependence of 
\vtwo on the shape of the nuclear overlap region. The $N_{\rm 
part}^{1/3}$ factor is introduced to characterize the length scale of 
nuclear overlap region and assumed to be proportional to the QGP length 
scale~\cite{FlowCuCuAuAu}. This suggests that the influence of the 
initial conditions on \vtwo coefficients, and thereby on QGP properties, 
are reasonably well encapsulated in the scaling factor $\ecc N_{\rm 
part}^{1/3}$. The scaling of \vtwo values with the shape and size of 
nuclear-overlap region can be explained by the hydrodynamic nature of the 
QGP at low values of specific-shear viscosity~\cite{hydrodynamic}.

The comparisons of elliptic-flow \vtwo and $\vtwo/\nq$ values obtained 
for \vphi mesons in 0\%--20\% and 20\%--60\% \cuau collisions and 
0\%--50\% \uu collisions to corresponding \vtwo and $\vtwo/\nq$ values 
for $\pi^\pm$ mesons and (anti)protons 
($(p+\bar{p})/2$)~\cite{FlowCuAu,v2_UU} are shown in 
Figs.~\ref{fig:Flow_CuAu_phi_pi0_prot_centr,fig:Flow_CuAu_phi_pi0_prot_periph,
fig:Flow_UU_phi_pi0_prot}, respectively. The scaling of light 
hadron $\vtwo$ with the number of valence quarks in the hadron \nq and 
transverse kinetic energy per valence quark $\kEt/\nq$ is observed. The 
\nq scaling can be explained via quark-coalescence models in which 
partons develop flow during the evolution of partonic matter and the 
hadron flow is the sum of collective flows of constituent 
partons~\cite{RecombinationFlow1,RecombinationFlow2}. A smaller 
rescattering cross section~\cite{Phi_cross_section} for \vphi mesons 
than for $\pi^\pm$ mesons and (anti)protons may also indicate that the 
elliptic flow develops prior to hadronization.


\begin{table*}[tbh]
\caption{\label{table:PYTHIA_parameters}
Parameters used in PYTHIA/Angantyr }
\begin{ruledtabular}
\begin{tabular}{ccc}
Parameter     & Value  &  Description                            \\ \hline
SoftQCD: all   & on     & All soft QCD processes                  \\
PDF: pSet      & 8      & CTEQ6l1 parton distribution function    \\
Multiparton Interactions: Kfactor  & 0.5 
& Multiplication factor for multiparton interaction              \\
\end{tabular}
\end{ruledtabular}
\end{table*}

\begin{table}[tbh]
\caption{\label{table:p_val_RAB_CuAu}
$p$-values estimated for the string melting version of AMPT and 
PYTHIA/Angantyr calculations of \vphi-meson \rab in different centrality 
classes of \cuau collisions at \sqsn = 200 GeV at midrapidity.}
\begin{ruledtabular}
\begin{tabular}{ccc}
Centrality       & \multicolumn{2}{c}{$p$ value} \\
                 & AMPT sm   & PYTHIA/Angantyr   \\ \hline
0\%--20\%      & 0.823  & 3.22$\times10^{-4}$    \\
20\%--40\%     & 0.712  & 5.79$\times10^{-5}$    \\
40\%--60\%     & 0.103  & 4.88$\times10^{-3}$    \\
60\%--80\%     & 0.671  & 0.455                  \\
\end{tabular}
\end{ruledtabular}
\end{table}

\begin{table}[tbh]
\caption{\label{table:p_val_v2}
$p$-values estimated for the iEBE-VISHNU and string melting
version of AMPT calculations of \vphi-meson \vtwo in different
centrality classes of \cuau at \sqsn = 200 GeV and \uu
collisions at \sqsn = 193 GeV.}
\begin{ruledtabular}
\begin{tabular}{cccc}
Collision &  Centrality      & \multicolumn{2}{c}{$p$ value}   \\
          &                  & iEBE-VISHNU & AMPT sm     \\ \hline
\cuau     & 0\%--20\%      & 0.788    & 0.287   \\
          & 20\%--40\%     & 0.985    & 0.927  \\
          & 40\%--60\%     & 0.998    & 0.878  \\
          & 20\%--60\%     & 0.905    & 0.513      \\
\uu       & 0\%--50\%      & 0.756    & 0.097   \\
\end{tabular}
\end{ruledtabular}
\end{table}

The elliptic-flow values measured for \vphi mesons are compared to the 
calculations of iEBE-VISHNU (2+1)D viscous-hydrodynamic model with 
specific viscosity $\et/s=1/(4\pi)$ and string-melting version of the 
AMPT model. The comparisons of measured \vphi-meson \vtwo values to AMPT 
and iEBE-VISHNU model predictions are shown in Fig.~\ref{fig:Flow_iEBE} 
for 0\%--20\%, 20\%--40\%, 40\%--60\%, and 20\%--60\% \cuau collisions 
and for 0\%--50\% \uu collisions. Table~\ref{table:p_val_v2} shows the 
$p$-values estimated for iEBE-VISHNU and AMPT model calculations of 
\vphi-meson \vtwo values in different centrality classes of \cuau and 
\uu collisions.

Elliptic flow for \vphi mesons estimated with the AMPT model are consistent 
within uncertainties with the \vphi-meson \vtwo values measured in \cuau 
collisions. The \vphi-meson \vtwo values in \uu collisions are 
under predicted by AMPT calculations, as shown in 
Fig.~\ref{fig:Flow_iEBE}(e). In contrast, calculations of 
iEBE-VISHNU (2+1)D viscous-hydrodynamic model with specific-shear 
viscosity $\et/s=1/(4\pi)$ reproduce \vphi-meson elliptic flow measured 
in both \cuau and \uu collisions with high precision.



\section{SUMMARY}

The PHENIX experiment has measured invariant transverse-momentum 
spectra, nuclear-modification factors, and elliptic flow for \vphi 
mesons in asymmetric \cuau collisions at \sqsntwo and in the largest 
collision system at RHIC, \uu at \sqsnuu at midrapidity \midrap via the 
kaon-decay channel. The comparisons of measured \vphi-meson \rab and 
\vtwo values to previously obtained PHENIX results and to model predictions 
have been provided.

It is found that features of \vphi-meson production measured in heavy 
ion collisions reported by the PHENIX experiment do not depend on the 
shape of the nuclear-overlap region. The obtained \vphi-meson 
$\langle{\rab}\rangle$ and $\vtwo/(\ecc N_{\rm part}^{1/3})$ values are 
consistent across \cucu, \cuau, \auau, and \uu collisions within 
uncertainties. The measured \vphi-meson production averaged over the 
azimuthal angle scales with the nuclear-overlap size. Elliptic flow for 
\vphi mesons scales with the second-order-participant eccentricity and 
the characteristic length of the nuclear-overlap area.

The \vphi-meson \rab values measured in \cuau and \uu collisions at 
moderate \pt are larger than \rab values of \pio and \et mesons. In both 
\cuau and \uu collisions, the \vphi-meson \vtwo values follow the 
patterns of {\pio}-meson and \prots \vtwo values when scaled with the 
number of valence quarks in the hadron \nq. Both of these observations 
at moderate \pt can be qualitatively explained by recombination of 
$s\bar{s}$ pairs in \vphi-meson production. The obtained \vphi-meson 
\rab and \vtwo values are quantitatively described by the AMPT and 
iEBE-VISHNU models, which include the coalescence mechanism.

At high \pt, yields of \vphi, \pio, and \et mesons are equally 
suppressed in \cuau and \uu collisions. This pattern is in agreement 
with expectations for in-medium energy loss of parent partons prior to 
their fragmentation. The high-\pt suppression scales with size of the 
nuclear-overlap region, which is assumed to be proportional to the QGP 
volume.

The scaling of hadronic elliptic flow with the number of valence quarks 
in the hadron \nq, the second-order-participant eccentricity \ecc, and 
the cube root of participant-nucleons number $N_{\rm part}^{1/3}$ can be 
explained by the hydrodynamic nature of the QGP. The measured \vtwo 
values for \vphi mesons are well described by (2+1)D viscous 
hydrodynamic model with specific shear viscosity $\et/s=1/(4\pi)$.




\begin{acknowledgments}

We thank the staff of the Collider-Accelerator and Physics
Departments at Brookhaven National Laboratory and the staff of
the other PHENIX participating institutions for their vital
contributions.  
We acknowledge support from the Office of Nuclear Physics in the
Office of Science of the Department of Energy,
the National Science Foundation,
Abilene Christian University Research Council,
Research Foundation of SUNY, and
Dean of the College of Arts and Sciences, Vanderbilt University
(U.S.A),
Ministry of Education, Culture, Sports, Science, and Technology
and the Japan Society for the Promotion of Science (Japan),
Natural Science Foundation of China (People's Republic of China),
Croatian Science Foundation and
Ministry of Science and Education (Croatia),
Ministry of Education, Youth and Sports (Czech Republic),
Centre National de la Recherche Scientifique, Commissariat
{\`a} l'{\'E}nergie Atomique, and Institut National de Physique
Nucl{\'e}aire et de Physique des Particules (France),
J. Bolyai Research Scholarship, EFOP, the New National Excellence
Program ({\'U}NKP), NKFIH, and OTKA (Hungary),
Department of Atomic Energy and Department of Science and Technology
(India),
Israel Science Foundation (Israel),
Basic Science Research and SRC(CENuM) Programs through NRF
funded by the Ministry of Education and the Ministry of
Science and ICT (Korea).
Ministry of Education and Science, Russian Academy of Sciences,
Federal Agency of Atomic Energy (Russia),
VR and Wallenberg Foundation (Sweden),
University of Zambia, the Government of the Republic of Zambia (Zambia),
the U.S. Civilian Research and Development Foundation for the
Independent States of the Former Soviet Union,
the Hungarian American Enterprise Scholarship Fund,
the US-Hungarian Fulbright Foundation,
and the US-Israel Binational Science Foundation.

\end{acknowledgments}


\begin{thebibliography}{60}%
\makeatletter
\providecommand \@ifxundefined [1]{%
 \@ifx{#1\undefined}
}%
\providecommand \@ifnum [1]{%
 \ifnum #1\expandafter \@firstoftwo
 \else \expandafter \@secondoftwo
 \fi
}%
\providecommand \@ifx [1]{%
 \ifx #1\expandafter \@firstoftwo
 \else \expandafter \@secondoftwo
 \fi
}%
\providecommand \natexlab [1]{#1}%
\providecommand \enquote  [1]{``#1''}%
\providecommand \bibnamefont  [1]{#1}%
\providecommand \bibfnamefont [1]{#1}%
\providecommand \citenamefont [1]{#1}%
\providecommand \href@noop [0]{\@secondoftwo}%
\providecommand \href [0]{\begingroup \@sanitize@url \@href}%
\providecommand \@href[1]{\@@startlink{#1}\@@href}%
\providecommand \@@href[1]{\endgroup#1\@@endlink}%
\providecommand \@sanitize@url [0]{\catcode `\\12\catcode `\$12\catcode
  `\&12\catcode `\#12\catcode `\^12\catcode `\_12\catcode `\%12\relax}%
\providecommand \@@startlink[1]{}%
\providecommand \@@endlink[0]{}%
\providecommand \url  [0]{\begingroup\@sanitize@url \@url }%
\providecommand \@url [1]{\endgroup\@href {#1}{\urlprefix }}%
\providecommand \urlprefix  [0]{URL }%
\providecommand \Eprint [0]{\href }%
\providecommand \doibase [0]{https://doi.org/}%
\providecommand \selectlanguage [0]{\@gobble}%
\providecommand \bibinfo  [0]{\@secondoftwo}%
\providecommand \bibfield  [0]{\@secondoftwo}%
\providecommand \translation [1]{[#1]}%
\providecommand \BibitemOpen [0]{}%
\providecommand \bibitemStop [0]{}%
\providecommand \bibitemNoStop [0]{.\EOS\space}%
\providecommand \EOS [0]{\spacefactor3000\relax}%
\providecommand \BibitemShut  [1]{\csname bibitem#1\endcsname}%
\let\auto@bib@innerbib\@empty
\bibitem [{\citenamefont {Harrison}\ \emph {et~al.}(2003)\citenamefont
  {Harrison}, \citenamefont {Ludlam},\ and\ \citenamefont {Ozaki}}]{RHIC}%
  \BibitemOpen
  \bibfield  {author} {\bibinfo {author} {\bibfnamefont {M.}~\bibnamefont
  {Harrison}}, \bibinfo {author} {\bibfnamefont {T.}~\bibnamefont {Ludlam}},\
  and\ \bibinfo {author} {\bibfnamefont {S.}~\bibnamefont {Ozaki}},\ }\bibfield
   {title} {\bibinfo {title} {{RHIC project overview}},\ }\href
  {https://doi.org/10.1016/S0168-9002(02)01937-X} {\bibfield  {journal}
  {\bibinfo  {journal} {Nucl. Instrum. Methods Phys. Res., Sec. A}\ }\textbf
  {\bibinfo {volume} {499}},\ \bibinfo {pages} {235} (\bibinfo {year}
  {2003})}\BibitemShut {NoStop}%
\bibitem [{\citenamefont {Adcox}\ \emph {et~al.}(2005)\citenamefont {Adcox}
  \emph {et~al.}}]{QGP}%
  \BibitemOpen
  \bibfield  {author} {\bibinfo {author} {\bibfnamefont {K.}~\bibnamefont
  {Adcox}} \emph {et~al.} (\bibinfo {collaboration} {PHENIX Collaboration}),\
  }\bibfield  {title} {\bibinfo {title} {{Formation of dense partonic matter in
  relativistic nucleus-nucleus collisions at RHIC: Experimental evaluation by
  the PHENIX collaboration}},\ }\href
  {https://doi.org/10.1016/j.nuclphysa.2005.03.086} {\bibfield  {journal}
  {\bibinfo  {journal} {Nucl. Phys. A}\ }\textbf {\bibinfo {volume} {757}},\
  \bibinfo {pages} {184} (\bibinfo {year} {2005})}\BibitemShut {NoStop}%
\bibitem [{\citenamefont {Arsene}\ \emph {et~al.}(2005)\citenamefont {Arsene}
  \emph {et~al.}}]{BRAHMS}%
  \BibitemOpen
  \bibfield  {author} {\bibinfo {author} {\bibfnamefont {I.}~\bibnamefont
  {Arsene}} \emph {et~al.} (\bibinfo {collaboration} {BRAHMS Collaboration}),\
  }\bibfield  {title} {\bibinfo {title} {{Quark gluon plasma and color glass
  condensate at RHIC? The Perspective from the BRAHMS experiment}},\ }\href
  {https://doi.org/10.1016/j.nuclphysa.2005.02.130} {\bibfield  {journal}
  {\bibinfo  {journal} {Nucl. Phys. A}\ }\textbf {\bibinfo {volume} {757}},\
  \bibinfo {pages} {1} (\bibinfo {year} {2005})}\BibitemShut {NoStop}%
\bibitem [{\citenamefont {Back}\ \emph {et~al.}(2005)\citenamefont {Back} \emph
  {et~al.}}]{PHOBOS}%
  \BibitemOpen
  \bibfield  {author} {\bibinfo {author} {\bibfnamefont {B.~B.}\ \bibnamefont
  {Back}} \emph {et~al.} (\bibinfo {collaboration} {PHOBOS Collaboration}),\
  }\bibfield  {title} {\bibinfo {title} {{The PHOBOS perspective on discoveries
  at RHIC}},\ }\href {https://doi.org/10.1016/j.nuclphysa.2005.03.084}
  {\bibfield  {journal} {\bibinfo  {journal} {Nucl. Phys. A}\ }\textbf
  {\bibinfo {volume} {757}},\ \bibinfo {pages} {28} (\bibinfo {year}
  {2005})}\BibitemShut {NoStop}%
\bibitem [{\citenamefont {Adams}\ \emph {et~al.}(2005)\citenamefont {Adams}
  \emph {et~al.}}]{STAR}%
  \BibitemOpen
  \bibfield  {author} {\bibinfo {author} {\bibfnamefont {J.}~\bibnamefont
  {Adams}} \emph {et~al.} (\bibinfo {collaboration} {STAR Collaboration}),\
  }\bibfield  {title} {\bibinfo {title} {{Experimental and theoretical
  challenges in the search for the quark gluon plasma: The STAR Collaboration's
  critical assessment of the evidence from RHIC collisions}},\ }\href
  {https://doi.org/10.1016/j.nuclphysa.2005.03.085} {\bibfield  {journal}
  {\bibinfo  {journal} {Nucl. Phys. A}\ }\textbf {\bibinfo {volume} {757}},\
  \bibinfo {pages} {102} (\bibinfo {year} {2005})}\BibitemShut {NoStop}%
\bibitem [{\citenamefont {Chatrchyan}\ \emph {et~al.}(2012)\citenamefont
  {Chatrchyan} \emph {et~al.}}]{LHC}%
  \BibitemOpen
  \bibfield  {author} {\bibinfo {author} {\bibfnamefont {S.}~\bibnamefont
  {Chatrchyan}} \emph {et~al.} (\bibinfo {collaboration} {CMS Collaboration}),\
  }\bibfield  {title} {\bibinfo {title} {{Study of high-pT charged particle
  suppression in PbPb compared to $pp$ collisions at $\sqrt{s_{NN}}=2.76$
  TeV}},\ }\href {https://doi.org/10.1140/epjc/s10052-012-1945-x} {\bibfield
  {journal} {\bibinfo  {journal} {Eur. Phys. J. C}\ }\textbf {\bibinfo {volume}
  {72}},\ \bibinfo {pages} {1945} (\bibinfo {year} {2012})}\BibitemShut
  {NoStop}%
\bibitem [{\citenamefont {Abelev}\ \emph {et~al.}(2013)\citenamefont {Abelev}
  \emph {et~al.}}]{ALICE}%
  \BibitemOpen
  \bibfield  {author} {\bibinfo {author} {\bibfnamefont {B.}~\bibnamefont
  {Abelev}} \emph {et~al.} (\bibinfo {collaboration} {ALICE Collaboration}),\
  }\bibfield  {title} {\bibinfo {title} {{Centrality Dependence of Charged
  Particle Production at Large Transverse Momentum in Pb--Pb Collisions at
  $\sqrt{s_{\rm{NN}}} = 2.76$ TeV}},\ }\href
  {https://doi.org/10.1016/j.physletb.2013.01.051} {\bibfield  {journal}
  {\bibinfo  {journal} {Phys. Lett. B}\ }\textbf {\bibinfo {volume} {720}},\
  \bibinfo {pages} {52} (\bibinfo {year} {2013})}\BibitemShut {NoStop}%
\bibitem [{\citenamefont {Aad}\ \emph {et~al.}(2013)\citenamefont {Aad} \emph
  {et~al.}}]{ATLAS}%
  \BibitemOpen
  \bibfield  {author} {\bibinfo {author} {\bibfnamefont {G.}~\bibnamefont
  {Aad}} \emph {et~al.} (\bibinfo {collaboration} {ATLAS Collaboration}),\
  }\bibfield  {title} {\bibinfo {title} {{Measurement of the jet radius and
  transverse momentum dependence of inclusive jet suppression in lead-lead
  collisions at $\sqrt{s_{NN}}$= 2.76 TeV with the ATLAS detector}},\ }\href
  {https://doi.org/10.1016/j.physletb.2013.01.024} {\bibfield  {journal}
  {\bibinfo  {journal} {Phys. Lett. B}\ }\textbf {\bibinfo {volume} {719}},\
  \bibinfo {pages} {220} (\bibinfo {year} {2013})}\BibitemShut {NoStop}%
\bibitem [{\citenamefont {Gunji}(2016)}]{ALICE1}%
  \BibitemOpen
  \bibfield  {author} {\bibinfo {author} {\bibfnamefont {T.}~\bibnamefont
  {Gunji}} (\bibinfo {collaboration} {ALICE Collaboration}),\ }\bibfield
  {title} {\bibinfo {title} {{Overview of recent ALICE results}},\ }\href
  {https://doi.org/10.1016/j.nuclphysa.2016.02.038} {\bibfield  {journal}
  {\bibinfo  {journal} {Nucl. Phys. A}\ }\textbf {\bibinfo {volume} {956}},\
  \bibinfo {pages} {11} (\bibinfo {year} {2016})}\BibitemShut {NoStop}%
\bibitem [{\citenamefont {Adcox}\ \emph
  {et~al.}(2003{\natexlab{a}})\citenamefont {Adcox} \emph
  {et~al.}}]{PHENIXoverview}%
  \BibitemOpen
  \bibfield  {author} {\bibinfo {author} {\bibfnamefont {K.}~\bibnamefont
  {Adcox}} \emph {et~al.} (\bibinfo {collaboration} {PHENIX Collaboration}),\
  }\bibfield  {title} {\bibinfo {title} {{PHENIX detector overview}},\ }\href
  {https://doi.org/10.1016/S0168-9002(02)01950-2} {\bibfield  {journal}
  {\bibinfo  {journal} {Nucl. Instrum. Methods Phys. Res., Sec. A}\ }\textbf
  {\bibinfo {volume} {499}},\ \bibinfo {pages} {469} (\bibinfo {year}
  {2003}{\natexlab{a}})}\BibitemShut {NoStop}%
\bibitem [{\citenamefont {Ethier}\ and\ \citenamefont {Nocera}(2020)}]{nPDFs}%
  \BibitemOpen
  \bibfield  {author} {\bibinfo {author} {\bibfnamefont {J.~J.}\ \bibnamefont
  {Ethier}}\ and\ \bibinfo {author} {\bibfnamefont {E.~R.}\ \bibnamefont
  {Nocera}},\ }\bibfield  {title} {\bibinfo {title} {{Parton Distributions in
  Nucleons and Nuclei}},\ }\href {https://doi.org/10.
  1146/annurev-nucl-011720-042725} {\bibfield  {journal} {\bibinfo  {journal}
  {Ann. Rev. Nucl. Part. Sci.}\ }\textbf {\bibinfo {volume} {70}},\ \bibinfo
  {pages} {43} (\bibinfo {year} {2020})}\BibitemShut {NoStop}%
\bibitem [{\citenamefont {Aidala}\ \emph {et~al.}(2018)\citenamefont {Aidala}
  \emph {et~al.}}]{Pi_eta_cuau}%
  \BibitemOpen
  \bibfield  {author} {\bibinfo {author} {\bibfnamefont {C.}~\bibnamefont
  {Aidala}} \emph {et~al.} (\bibinfo {collaboration} {PHENIX Collaboration}),\
  }\bibfield  {title} {\bibinfo {title} {{Production of
  ${\ensuremath{\pi}}^{0}$ and $\ensuremath{\eta}$ mesons in Cu$+$Au collisions
  at $\sqrt{s_{_{NN}}}=200$ GeV}},\ }\href
  {https://doi.org/10.1103/PhysRevC.98.054903} {\bibfield  {journal} {\bibinfo
  {journal} {Phys. Rev. C}\ }\textbf {\bibinfo {volume} {98}},\ \bibinfo
  {pages} {054903} (\bibinfo {year} {2018})}\BibitemShut {NoStop}%
\bibitem [{\citenamefont {d'Enterria}(2010)}]{JetQuenching}%
  \BibitemOpen
  \bibfield  {author} {\bibinfo {author} {\bibfnamefont {D.}~\bibnamefont
  {d'Enterria}},\ }\bibfield  {title} {\bibinfo {title} {{Jet quenching}},\
  }\href {https://doi.org/10.1007/978-3-642-01539-7_16} {\bibfield  {journal}
  {\bibinfo  {journal} {Landolt-Bornstein}\ }\textbf {\bibinfo {volume} {23}},\
  \bibinfo {pages} {471} (\bibinfo {year} {2010})}\BibitemShut {NoStop}%
\bibitem [{\citenamefont {Koch}\ \emph {et~al.}(1986)\citenamefont {Koch},
  \citenamefont {Muller},\ and\ \citenamefont
  {Rafelski}}]{StrangenessEnhancenment}%
  \BibitemOpen
  \bibfield  {author} {\bibinfo {author} {\bibfnamefont {P.}~\bibnamefont
  {Koch}}, \bibinfo {author} {\bibfnamefont {B.}~\bibnamefont {Muller}},\ and\
  \bibinfo {author} {\bibfnamefont {J.}~\bibnamefont {Rafelski}},\ }\bibfield
  {title} {\bibinfo {title} {{Strangeness in Relativistic Heavy Ion
  Collisions}},\ }\href {https://doi.org/10.1016/0370-1573(86)90096-7}
  {\bibfield  {journal} {\bibinfo  {journal} {Phys. Rept.}\ }\textbf {\bibinfo
  {volume} {142}},\ \bibinfo {pages} {167} (\bibinfo {year}
  {1986})}\BibitemShut {NoStop}%
\bibitem [{\citenamefont {Greco}\ \emph {et~al.}(2003)\citenamefont {Greco},
  \citenamefont {Ko},\ and\ \citenamefont {L\'evai}}]{Recombination1}%
  \BibitemOpen
  \bibfield  {author} {\bibinfo {author} {\bibfnamefont {V.}~\bibnamefont
  {Greco}}, \bibinfo {author} {\bibfnamefont {C.~M.}\ \bibnamefont {Ko}},\ and\
  \bibinfo {author} {\bibfnamefont {P.}~\bibnamefont {L\'evai}},\ }\bibfield
  {title} {\bibinfo {title} {{Parton Coalescence and the Antiproton/Pion
  Anomaly at RHIC}},\ }\href {https://doi.org/10.1103/PhysRevLett.90.202302}
  {\bibfield  {journal} {\bibinfo  {journal} {Phys. Rev. Lett.}\ }\textbf
  {\bibinfo {volume} {90}},\ \bibinfo {pages} {202302} (\bibinfo {year}
  {2003})}\BibitemShut {NoStop}%
\bibitem [{\citenamefont {Hwa}\ and\ \citenamefont
  {Yang}(2003)}]{Recombination2}%
  \BibitemOpen
  \bibfield  {author} {\bibinfo {author} {\bibfnamefont {R.~C.}\ \bibnamefont
  {Hwa}}\ and\ \bibinfo {author} {\bibfnamefont {C.~B.}\ \bibnamefont {Yang}},\
  }\bibfield  {title} {\bibinfo {title} {{Scaling behavior at high $p_T$ and
  the $p/\pi$ ratio}},\ }\href {https://doi.org/10.1103/PhysRevC.67.034902}
  {\bibfield  {journal} {\bibinfo  {journal} {Phys. Rev. C}\ }\textbf {\bibinfo
  {volume} {67}},\ \bibinfo {pages} {034902} (\bibinfo {year}
  {2003})}\BibitemShut {NoStop}%
\bibitem [{\citenamefont {Hwa}\ and\ \citenamefont
  {Yang}(2007)}]{Recombination_phi}%
  \BibitemOpen
  \bibfield  {author} {\bibinfo {author} {\bibfnamefont {R.~C.}\ \bibnamefont
  {Hwa}}\ and\ \bibinfo {author} {\bibfnamefont {C.~B.}\ \bibnamefont {Yang}},\
  }\bibfield  {title} {\bibinfo {title} {{Production of strange particles at
  intermediate $p_T$ in central Au$+$Au collisions at high energies}},\ }\href
  {https://doi.org/10.1103/physrevc.75.054904} {\bibfield  {journal} {\bibinfo
  {journal} {Phys. Rev. C}\ }\textbf {\bibinfo {volume} {75}},\ \bibinfo
  {pages} {054904} (\bibinfo {year} {2007})}\BibitemShut {NoStop}%
\bibitem [{\citenamefont {Borghini}(2005)}]{Flow_meth}%
  \BibitemOpen
  \bibfield  {author} {\bibinfo {author} {\bibfnamefont {N.}~\bibnamefont
  {Borghini}},\ }\bibfield  {title} {\bibinfo {title} {{Characterization and
  analysis of azimuthally sensitive correlations}},\ }\href
  {https://doi.org/10.1088/0954-3899/31/4/002} {\bibfield  {journal} {\bibinfo
  {journal} {J. Phys. G}\ }\textbf {\bibinfo {volume} {31}},\ \bibinfo {pages}
  {S15} (\bibinfo {year} {2005})}\BibitemShut {NoStop}%
\bibitem [{\citenamefont {Adare}\ \emph {et~al.}(2015)\citenamefont {Adare}
  \emph {et~al.}}]{FlowCuCuAuAu}%
  \BibitemOpen
  \bibfield  {author} {\bibinfo {author} {\bibfnamefont {A.}~\bibnamefont
  {Adare}} \emph {et~al.} (\bibinfo {collaboration} {PHENIX Collaboration}),\
  }\bibfield  {title} {\bibinfo {title} {{Systematic Study of Azimuthal
  Anisotropy in Cu$+$Cu and Au$+$Au Collisions at $\sqrt{s_{NN}} = 62. 4$ and
  200 GeV}},\ }\href {https://doi.org/10. 1103/PhysRevC.92.034913} {\bibfield
  {journal} {\bibinfo  {journal} {Phys. Rev. C}\ }\textbf {\bibinfo {volume}
  {92}},\ \bibinfo {pages} {034913} (\bibinfo {year} {2015})}\BibitemShut
  {NoStop}%
\bibitem [{\citenamefont {Adler}\ \emph
  {et~al.}(2005{\natexlab{a}})\citenamefont {Adler} \emph {et~al.}}]{Flow}%
  \BibitemOpen
  \bibfield  {author} {\bibinfo {author} {\bibfnamefont {S.~S.}\ \bibnamefont
  {Adler}} \emph {et~al.} (\bibinfo {collaboration} {PHENIX Collaboration}),\
  }\bibfield  {title} {\bibinfo {title} {{Saturation of Azimuthal Anisotropy in
  Au$+$Au Collisions at $\sqrt{s_{_{NN}}}=62$--200 GeV}},\ }\href
  {https://doi.org/10.1103/PhysRevLett.94.232302} {\bibfield  {journal}
  {\bibinfo  {journal} {Phys. Rev. Lett.}\ }\textbf {\bibinfo {volume} {94}},\
  \bibinfo {pages} {232302} (\bibinfo {year} {2005}{\natexlab{a}})}\BibitemShut
  {NoStop}%
\bibitem [{\citenamefont {Heinz}\ and\ \citenamefont
  {Snellings}(2013)}]{hydrodynamic}%
  \BibitemOpen
  \bibfield  {author} {\bibinfo {author} {\bibfnamefont {U.}~\bibnamefont
  {Heinz}}\ and\ \bibinfo {author} {\bibfnamefont {R.}~\bibnamefont
  {Snellings}},\ }\bibfield  {title} {\bibinfo {title} {{Collective Flow and
  Viscosity in Relativistic Heavy-Ion Collisions}},\ }\href
  {https://doi.org/10.1146/annurev-nucl-102212-170540} {\bibfield  {journal}
  {\bibinfo  {journal} {Ann. Rev. Nucl. Part. Sci.}\ }\textbf {\bibinfo
  {volume} {63}},\ \bibinfo {pages} {123} (\bibinfo {year} {2013})}\BibitemShut
  {NoStop}%
\bibitem [{\citenamefont {Shor}(1985)}]{PhiQGP}%
  \BibitemOpen
  \bibfield  {author} {\bibinfo {author} {\bibfnamefont {A.}~\bibnamefont
  {Shor}},\ }\bibfield  {title} {\bibinfo {title} {{$\ensuremath{\phi}$-meson
  production as a probe of the Quark-Gluon Plasma}},\ }\href
  {https://doi.org/10.1103/PhysRevLett.54.1122} {\bibfield  {journal} {\bibinfo
   {journal} {Phys. Rev. Lett.}\ }\textbf {\bibinfo {volume} {54}},\ \bibinfo
  {pages} {1122} (\bibinfo {year} {1985})}\BibitemShut {NoStop}%
\bibitem [{\citenamefont {Kolb}(2004)}]{QGP_expansion_rates}%
  \BibitemOpen
  \bibfield  {author} {\bibinfo {author} {\bibfnamefont {P.~F.}\ \bibnamefont
  {Kolb}},\ }\bibfield  {title} {\bibinfo {title} {{Expansion rates at RHIC}},\
  }\href {https://doi.org/10.1556/APH.21.2004.2-4.22} {\bibfield  {journal}
  {\bibinfo  {journal} {Acta Phys. Hung. A}\ }\textbf {\bibinfo {volume}
  {21}},\ \bibinfo {pages} {243} (\bibinfo {year} {2004})}\BibitemShut
  {NoStop}%
\bibitem [{\citenamefont {Lin}\ \emph {et~al.}(2005)\citenamefont {Lin},
  \citenamefont {Ko}, \citenamefont {Li}, \citenamefont {Zhang},\ and\
  \citenamefont {Pal}}]{AMPT}%
  \BibitemOpen
  \bibfield  {author} {\bibinfo {author} {\bibfnamefont {Z.-W.}\ \bibnamefont
  {Lin}}, \bibinfo {author} {\bibfnamefont {C.~M.}\ \bibnamefont {Ko}},
  \bibinfo {author} {\bibfnamefont {B.-A.}\ \bibnamefont {Li}}, \bibinfo
  {author} {\bibfnamefont {B.}~\bibnamefont {Zhang}},\ and\ \bibinfo {author}
  {\bibfnamefont {S.}~\bibnamefont {Pal}},\ }\bibfield  {title} {\bibinfo
  {title} {{A Multi-phase transport model for relativistic heavy ion
  collisions}},\ }\href {https://doi.org/10.1103/PhysRevC.72.064901} {\bibfield
   {journal} {\bibinfo  {journal} {Phys. Rev. C}\ }\textbf {\bibinfo {volume}
  {72}},\ \bibinfo {pages} {064901} (\bibinfo {year} {2005})}\BibitemShut
  {NoStop}%
\bibitem [{\citenamefont {Shen}\ \emph {et~al.}(2016)\citenamefont {Shen},
  \citenamefont {Qiu}, \citenamefont {Song}, \citenamefont {Bernhard},
  \citenamefont {Bass},\ and\ \citenamefont {Heinz}}]{iebevishnu}%
  \BibitemOpen
  \bibfield  {author} {\bibinfo {author} {\bibfnamefont {C.}~\bibnamefont
  {Shen}}, \bibinfo {author} {\bibfnamefont {Z.}~\bibnamefont {Qiu}}, \bibinfo
  {author} {\bibfnamefont {H.}~\bibnamefont {Song}}, \bibinfo {author}
  {\bibfnamefont {J.}~\bibnamefont {Bernhard}}, \bibinfo {author}
  {\bibfnamefont {S.}~\bibnamefont {Bass}},\ and\ \bibinfo {author}
  {\bibfnamefont {U.}~\bibnamefont {Heinz}},\ }\bibfield  {title} {\bibinfo
  {title} {{The iEBE-VISHNU code package for relativistic heavy-ion
  collisions}},\ }\href {https://doi.org/10.1016/j.cpc.2015.08.039} {\bibfield
  {journal} {\bibinfo  {journal} {Comput. Phys. Commun.}\ }\textbf {\bibinfo
  {volume} {199}},\ \bibinfo {pages} {61} (\bibinfo {year} {2016})}\BibitemShut
  {NoStop}%
\bibitem [{\citenamefont {da~Silva}\ \emph {et~al.}(2020)\citenamefont
  {da~Silva}, \citenamefont {Serenone}, \citenamefont {Chinellato},
  \citenamefont {Takahashi},\ and\ \citenamefont {Bierlich}}]{PYTHIA_ANGANTYR}%
  \BibitemOpen
  \bibfield  {author} {\bibinfo {author} {\bibfnamefont {A.~V.}\ \bibnamefont
  {da~Silva}}, \bibinfo {author} {\bibfnamefont {W.~M.}\ \bibnamefont
  {Serenone}}, \bibinfo {author} {\bibfnamefont {D.~D.}\ \bibnamefont
  {Chinellato}}, \bibinfo {author} {\bibfnamefont {J.}~\bibnamefont
  {Takahashi}},\ and\ \bibinfo {author} {\bibfnamefont {C.}~\bibnamefont
  {Bierlich}},\ }\href@noop {} {\bibinfo {title} {{Studies of heavy-ion
  collisions using PYTHIA Angantyr and UrQMD}}} (\bibinfo {year} {2020}),\
  \bibinfo {note} {{arXiv:2002.10236}}\BibitemShut {NoStop}%
\bibitem [{\citenamefont {Allen}\ \emph {et~al.}(2003)\citenamefont {Allen}
  \emph {et~al.}}]{BBC}%
  \BibitemOpen
  \bibfield  {author} {\bibinfo {author} {\bibfnamefont {M.}~\bibnamefont
  {Allen}} \emph {et~al.} (\bibinfo {collaboration} {PHENIX Collaboration}),\
  }\bibfield  {title} {\bibinfo {title} {{PHENIX inner detectors}},\ }\href
  {https://doi.org/https://doi.org/10.1016/S0168-9002(02)01956-3} {\bibfield
  {journal} {\bibinfo  {journal} {Nucl. Instrum. Methods Phys. Res., Sec. A}\
  }\textbf {\bibinfo {volume} {499}},\ \bibinfo {pages} {549} (\bibinfo {year}
  {2003})}\BibitemShut {NoStop}%
\bibitem [{\citenamefont {Adare}\ \emph {et~al.}(2014)\citenamefont {Adare}
  \emph {et~al.}}]{Centrality}%
  \BibitemOpen
  \bibfield  {author} {\bibinfo {author} {\bibfnamefont {A.}~\bibnamefont
  {Adare}} \emph {et~al.} (\bibinfo {collaboration} {PHENIX Collaboration}),\
  }\bibfield  {title} {\bibinfo {title} {{Centrality categorization for
  $R_{p(d)+A}$ in high-energy collisions}},\ }\href {https://doi.org/10.
  1103/PhysRevC.90.034902} {\bibfield  {journal} {\bibinfo  {journal} {Phys.
  Rev. C}\ }\textbf {\bibinfo {volume} {90}},\ \bibinfo {pages} {034902}
  (\bibinfo {year} {2014})}\BibitemShut {NoStop}%
\bibitem [{\citenamefont {Miller}\ \emph {et~al.}(2007)\citenamefont {Miller},
  \citenamefont {Reygers}, \citenamefont {Sanders},\ and\ \citenamefont
  {Steinberg}}]{Glauber}%
  \BibitemOpen
  \bibfield  {author} {\bibinfo {author} {\bibfnamefont {M.~L.}\ \bibnamefont
  {Miller}}, \bibinfo {author} {\bibfnamefont {K.}~\bibnamefont {Reygers}},
  \bibinfo {author} {\bibfnamefont {S.~J.}\ \bibnamefont {Sanders}},\ and\
  \bibinfo {author} {\bibfnamefont {P.}~\bibnamefont {Steinberg}},\ }\bibfield
  {title} {\bibinfo {title} {{Glauber Modeling in High-Energy Nuclear
  Collisions}},\ }\href {https://doi.org/10.
  1146/annurev.nucl.57.090506.123020} {\bibfield  {journal} {\bibinfo
  {journal} {Ann. Rev. Nucl. Part. Sci.}\ }\textbf {\bibinfo {volume} {57}},\
  \bibinfo {pages} {205} (\bibinfo {year} {2007})}\BibitemShut {NoStop}%
\bibitem [{\citenamefont {Masui}\ \emph {et~al.}(2009)\citenamefont {Masui},
  \citenamefont {Mohanty},\ and\ \citenamefont {Xu}}]{Glauber1}%
  \BibitemOpen
  \bibfield  {author} {\bibinfo {author} {\bibfnamefont {H.}~\bibnamefont
  {Masui}}, \bibinfo {author} {\bibfnamefont {B.}~\bibnamefont {Mohanty}},\
  and\ \bibinfo {author} {\bibfnamefont {N.}~\bibnamefont {Xu}},\ }\bibfield
  {title} {\bibinfo {title} {{Predictions of Elliptic flow and nuclear
  modification factor from 200 GeV U$+$U collisions at RHIC}},\ }\href
  {https://doi.org/https://doi.org/10.1016/j.physletb.2009.08.025} {\bibfield
  {journal} {\bibinfo  {journal} {Phys. Lett. B}\ }\textbf {\bibinfo {volume}
  {679}},\ \bibinfo {pages} {440} (\bibinfo {year} {2009})}\BibitemShut
  {NoStop}%
\bibitem [{\citenamefont {Shou}\ \emph {et~al.}(2015)\citenamefont {Shou},
  \citenamefont {Ma}, \citenamefont {Sorensen}, \citenamefont {Tang},
  \citenamefont {Videbaek},\ and\ \citenamefont {Wang}}]{Glauber2}%
  \BibitemOpen
  \bibfield  {author} {\bibinfo {author} {\bibfnamefont {Q.~Y.}\ \bibnamefont
  {Shou}}, \bibinfo {author} {\bibfnamefont {Y.~G.}\ \bibnamefont {Ma}},
  \bibinfo {author} {\bibfnamefont {P.}~\bibnamefont {Sorensen}}, \bibinfo
  {author} {\bibfnamefont {A.~H.}\ \bibnamefont {Tang}}, \bibinfo {author}
  {\bibfnamefont {F.}~\bibnamefont {Videbaek}},\ and\ \bibinfo {author}
  {\bibfnamefont {H.}~\bibnamefont {Wang}},\ }\bibfield  {title} {\bibinfo
  {title} {{Parameterization of deformed nuclei for {G}lauber modeling in
  relativistic heavy ion collisions}},\ }\href
  {https://doi.org/https://doi.org/10.1016/j.physletb.2015.07.078} {\bibfield
  {journal} {\bibinfo  {journal} {Phys. Lett. B}\ }\textbf {\bibinfo {volume}
  {749}},\ \bibinfo {pages} {215} (\bibinfo {year} {2015})}\BibitemShut
  {NoStop}%
\bibitem [{\citenamefont {Aidala}\ \emph {et~al.}(2014)\citenamefont {Aidala}
  \emph {et~al.}}]{FVTX}%
  \BibitemOpen
  \bibfield  {author} {\bibinfo {author} {\bibfnamefont {C.}~\bibnamefont
  {Aidala}} \emph {et~al.},\ }\bibfield  {title} {\bibinfo {title} {{The
  {PHENIX} Forward Silicon Vertex Detector}},\ }\href
  {https://doi.org/10.1016/j.nima.2014.04.017} {\bibfield  {journal} {\bibinfo
  {journal} {Nucl. Instrum. Methods Phys. Res., Sec. A}\ }\textbf {\bibinfo
  {volume} {755}},\ \bibinfo {pages} {44} (\bibinfo {year} {2014})}\BibitemShut
  {NoStop}%
\bibitem [{\citenamefont {Chiu}(2007)}]{MPC}%
  \BibitemOpen
  \bibfield  {author} {\bibinfo {author} {\bibfnamefont {M.}~\bibnamefont
  {Chiu}} (\bibinfo {collaboration} {PHENIX Collaboration}),\ }\bibfield
  {title} {\bibinfo {title} {{Single spin transverse asymmetries of neutral
  pions at forward rapidities in $\sqrt{s}=200$ = 62. 4 {GeV} polarized proton
  collisions in {PHENIX Collaboration}}},\ }\href {https://doi.org/10.
  1063/1.2750838} {\bibfield  {journal} {\bibinfo  {journal} {AIP Conf. Proc.}\
  }\textbf {\bibinfo {volume} {915}},\ \bibinfo {pages} {539} (\bibinfo {year}
  {2007})}\BibitemShut {NoStop}%
\bibitem [{\citenamefont {Poskanzer}\ and\ \citenamefont
  {Voloshin}(1998)}]{MethodsQ}%
  \BibitemOpen
  \bibfield  {author} {\bibinfo {author} {\bibfnamefont {A.~M.}\ \bibnamefont
  {Poskanzer}}\ and\ \bibinfo {author} {\bibfnamefont {S.~A.}\ \bibnamefont
  {Voloshin}},\ }\bibfield  {title} {\bibinfo {title} {{Methods for analyzing
  anisotropic flow in relativistic nuclear collisions}},\ }\href
  {https://doi.org/10.1103/PhysRevC.58.1671} {\bibfield  {journal} {\bibinfo
  {journal} {Phys. Rev. C}\ }\textbf {\bibinfo {volume} {58}},\ \bibinfo
  {pages} {1671} (\bibinfo {year} {1998})}\BibitemShut {NoStop}%
\bibitem [{\citenamefont {Barrette}\ \emph {et~al.}(1997)\citenamefont
  {Barrette} \emph {et~al.}}]{Flattening}%
  \BibitemOpen
  \bibfield  {author} {\bibinfo {author} {\bibfnamefont {J.}~\bibnamefont
  {Barrette}} \emph {et~al.} (\bibinfo {collaboration} {E877 Collaboration}),\
  }\bibfield  {title} {\bibinfo {title} {{Proton and pion production relative
  to the reaction plane in Au+Au collisions at AGS energies}},\ }\href
  {https://doi.org/10.1103/PhysRevC.56.3254} {\bibfield  {journal} {\bibinfo
  {journal} {Phys. Rev. C}\ }\textbf {\bibinfo {volume} {56}},\ \bibinfo
  {pages} {3254} (\bibinfo {year} {1997})}\BibitemShut {NoStop}%
\bibitem [{\citenamefont {Zyla}\ \emph {et~al.}(2020)\citenamefont {Zyla} \emph
  {et~al.}}]{RevPartPhys}%
  \BibitemOpen
  \bibfield  {author} {\bibinfo {author} {\bibfnamefont {P.~A.}\ \bibnamefont
  {Zyla}} \emph {et~al.} (\bibinfo {collaboration} {Particle Data Group}),\
  }\bibfield  {title} {\bibinfo {title} {{Rev. of Particle Phys.}},\ }\href
  {https://doi.org/10.1093/ptep/ptaa104} {\bibfield  {journal} {\bibinfo
  {journal} {Prog. Theor. Exp. Phys.}\ }\textbf {\bibinfo {volume} {2020}},\
  \bibinfo {pages} {083C01} (\bibinfo {year} {2020})}\BibitemShut {NoStop}%
\bibitem [{\citenamefont {Adler}\ \emph
  {et~al.}(2005{\natexlab{b}})\citenamefont {Adler} \emph {et~al.}}]{phi_AuAu}%
  \BibitemOpen
  \bibfield  {author} {\bibinfo {author} {\bibfnamefont {S.~S.}\ \bibnamefont
  {Adler}} \emph {et~al.} (\bibinfo {collaboration} {PHENIX Collaboration}),\
  }\bibfield  {title} {\bibinfo {title} {{Production of $\phi$ mesons at
  midrapidity in $\sqrt{s_{_{NN}}}=200$ GeV Au$+$Au collisions at RHIC}},\
  }\href@noop {} {\bibfield  {journal} {\bibinfo  {journal} {Phys. Rev. C}\
  }\textbf {\bibinfo {volume} {72}},\ \bibinfo {pages} {014903} (\bibinfo
  {year} {2005}{\natexlab{b}})}\BibitemShut {NoStop}%
\bibitem [{\citenamefont {Adare}\ \emph
  {et~al.}(2011{\natexlab{a}})\citenamefont {Adare} \emph
  {et~al.}}]{PhidAuCuCuAuAu}%
  \BibitemOpen
  \bibfield  {author} {\bibinfo {author} {\bibfnamefont {A.}~\bibnamefont
  {Adare}} \emph {et~al.} (\bibinfo {collaboration} {PHENIX Collaboration}),\
  }\bibfield  {title} {\bibinfo {title} {{Nuclear modification factors of
  $\phi$ mesons in $d$$+$Au, Cu$+$Cu and Au$+$Au collisions at
  $\sqrt{s_{_{NN}}}=200$ GeV}},\ }\href
  {https://doi.org/10.1103/PhysRevC.83.024909} {\bibfield  {journal} {\bibinfo
  {journal} {Phys. Rev. C}\ }\textbf {\bibinfo {volume} {83}},\ \bibinfo
  {pages} {024909} (\bibinfo {year} {2011}{\natexlab{a}})}\BibitemShut
  {NoStop}%
\bibitem [{\citenamefont {Afanasiev}\ \emph {et~al.}(2007)\citenamefont
  {Afanasiev} \emph {et~al.}}]{Flow_phi}%
  \BibitemOpen
  \bibfield  {author} {\bibinfo {author} {\bibfnamefont {S.}~\bibnamefont
  {Afanasiev}} \emph {et~al.} (\bibinfo {collaboration} {PHENIX
  Collaboration}),\ }\bibfield  {title} {\bibinfo {title} {{Elliptic flow for
  \ensuremath{\phi} mesons and (anti)deuterons in {A}u + {A}u collisions at
  $\sqrt {{s}_{_{NN}}} $ = 200 {G}e{V}}},\ }\href {https://doi.org/10.
  1103/PhysRevLett.99.052301} {\bibfield  {journal} {\bibinfo  {journal} {Phys.
  Rev. Lett.}\ }\textbf {\bibinfo {volume} {99}},\ \bibinfo {pages} {052301}
  (\bibinfo {year} {2007})}\BibitemShut {NoStop}%
\bibitem [{\citenamefont {Afanasiev}\ \emph {et~al.}(2009)\citenamefont
  {Afanasiev} \emph {et~al.}}]{Flow_pi0}%
  \BibitemOpen
  \bibfield  {author} {\bibinfo {author} {\bibfnamefont {S.}~\bibnamefont
  {Afanasiev}} \emph {et~al.} (\bibinfo {collaboration} {PHENIX
  Collaboration}),\ }\bibfield  {title} {\bibinfo {title} {{High-$p_T$ $\pi^0$
  Production with Respect to the Reaction Plane in Au$+$Au Collisions at
  $\sqrt{s_{_{NN}}}=200$ GeV}},\ }\href@noop {} {\bibfield  {journal} {\bibinfo
   {journal} {Phys. Rev. C}\ }\textbf {\bibinfo {volume} {80}},\ \bibinfo
  {pages} {054907} (\bibinfo {year} {2009})}\BibitemShut {NoStop}%
\bibitem [{\citenamefont {Adare}\ \emph {et~al.}(2013)\citenamefont {Adare}
  \emph {et~al.}}]{Flow_pi0_eta}%
  \BibitemOpen
  \bibfield  {author} {\bibinfo {author} {\bibfnamefont {A.}~\bibnamefont
  {Adare}} \emph {et~al.} (\bibinfo {collaboration} {PHENIX Collaboration}),\
  }\bibfield  {title} {\bibinfo {title} {{Azimuthal anisotropy of
  ${\ensuremath{\pi}}^{0}$ and $\ensuremath{\eta}$ mesons in Au$+$Au collisions
  at $\sqrt{s_{_{NN}}}=200$ {G}e{V}}},\ }\href
  {https://doi.org/10.1103/PhysRevC.88.064910} {\bibfield  {journal} {\bibinfo
  {journal} {Phys. Rev. C}\ }\textbf {\bibinfo {volume} {88}},\ \bibinfo
  {pages} {064910} (\bibinfo {year} {2013})}\BibitemShut {NoStop}%
\bibitem [{\citenamefont {Adcox}\ \emph
  {et~al.}(2003{\natexlab{b}})\citenamefont {Adcox} \emph
  {et~al.}}]{TrackingSystem}%
  \BibitemOpen
  \bibfield  {author} {\bibinfo {author} {\bibfnamefont {K.}~\bibnamefont
  {Adcox}} \emph {et~al.},\ }\bibfield  {title} {\bibinfo {title} {{PHENIX
  central arm tracking detectors}},\ }\href
  {https://doi.org/https://doi.org/10.1016/S0168-9002(02)01952-6} {\bibfield
  {journal} {\bibinfo  {journal} {Nucl. Instrum. Methods Phys. Res., Sec. A}\
  }\textbf {\bibinfo {volume} {499}},\ \bibinfo {pages} {489} (\bibinfo {year}
  {2003}{\natexlab{b}})}\BibitemShut {NoStop}%
\bibitem [{\citenamefont {Carl\'en}\ \emph {et~al.}(1999)\citenamefont
  {Carl\'en} \emph {et~al.}}]{TOF}%
  \BibitemOpen
  \bibfield  {author} {\bibinfo {author} {\bibfnamefont {L.}~\bibnamefont
  {Carl\'en}} \emph {et~al.},\ }\bibfield  {title} {\bibinfo {title} {{A
  large-acceptance spectrometer for tracking in a high multiplicity
  environment, based on space point measurements and high resolution
  time-of-flight}},\ }\href {https://doi.org/10.1016/S0168-9002(99)00261-2}
  {\bibfield  {journal} {\bibinfo  {journal} {Nucl. Instrum. Methods Phys.
  Res., Sec. A}\ }\textbf {\bibinfo {volume} {431}},\ \bibinfo {pages} {123}
  (\bibinfo {year} {1999})}\BibitemShut {NoStop}%
\bibitem [{\citenamefont {Aizawa}\ \emph {et~al.}(2003)\citenamefont {Aizawa}
  \emph {et~al.}}]{TOF2}%
  \BibitemOpen
  \bibfield  {author} {\bibinfo {author} {\bibfnamefont {M.}~\bibnamefont
  {Aizawa}} \emph {et~al.} (\bibinfo {collaboration} {PHENIX Collaboration}),\
  }\bibfield  {title} {\bibinfo {title} {{PHENIX central arm particle ID
  detectors}},\ }\href {https://doi.org/10.1016/S0168-9002(02)01953-8}
  {\bibfield  {journal} {\bibinfo  {journal} {Nucl. Instrum. Methods Phys.
  Res., Sec. A}\ }\textbf {\bibinfo {volume} {499}},\ \bibinfo {pages} {508}
  (\bibinfo {year} {2003})}\BibitemShut {NoStop}%
\bibitem [{\citenamefont {Acharya}\ \emph {et~al.}(2022)\citenamefont {Acharya}
  \emph {et~al.}}]{Phi_raw_yields}%
  \BibitemOpen
  \bibfield  {author} {\bibinfo {author} {\bibfnamefont {U.}~\bibnamefont
  {Acharya}} \emph {et~al.} (\bibinfo {collaboration} {PHENIX Collaboration}),\
  }\bibfield  {title} {\bibinfo {title} {{Study of $\phi$-meson production in
  $p$$+$Al, $p$$+$Au, $d$$+$Au, and $^3$He$+$Au collisions at
  $\sqrt{s_{_{NN}}}=200$ GeV}},\ }\href
  {https://doi.org/10.1103/PhysRevC.106.014908} {\bibfield  {journal} {\bibinfo
   {journal} {Phys. Rev. C}\ }\textbf {\bibinfo {volume} {106}},\ \bibinfo
  {pages} {014908} (\bibinfo {year} {2022})}\BibitemShut {NoStop}%
\bibitem [{\citenamefont {Brun}\ \emph {et~al.}(1987)\citenamefont {Brun},
  \citenamefont {Bruyant}, \citenamefont {Maire}, \citenamefont {McPherson},\
  and\ \citenamefont {Zanarini}}]{GEANT}%
  \BibitemOpen
  \bibfield  {author} {\bibinfo {author} {\bibfnamefont {R.}~\bibnamefont
  {Brun}}, \bibinfo {author} {\bibfnamefont {F.}~\bibnamefont {Bruyant}},
  \bibinfo {author} {\bibfnamefont {M.}~\bibnamefont {Maire}}, \bibinfo
  {author} {\bibfnamefont {A.~C.}\ \bibnamefont {McPherson}},\ and\ \bibinfo
  {author} {\bibfnamefont {P.}~\bibnamefont {Zanarini}},\ }\href@noop {}
  {\bibinfo {title} {{GEANT3}}} (\bibinfo {year} {1987}),\ \bibinfo {note}
  {https://github.com/vmc-project/geant3}\BibitemShut {NoStop}%
\bibitem [{\citenamefont {Mitrankova}\ \emph {et~al.}(2021)\citenamefont
  {Mitrankova}, \citenamefont {Berdnikov}, \citenamefont {Berdnikov},
  \citenamefont {Kotov},\ and\ \citenamefont {Mitrankov}}]{Mitrankova_2021}%
  \BibitemOpen
  \bibfield  {author} {\bibinfo {author} {\bibfnamefont {M.~M.}\ \bibnamefont
  {Mitrankova}}, \bibinfo {author} {\bibfnamefont {Y.~A.}\ \bibnamefont
  {Berdnikov}}, \bibinfo {author} {\bibfnamefont {A.~Y.}\ \bibnamefont
  {Berdnikov}}, \bibinfo {author} {\bibfnamefont {D.~O.}\ \bibnamefont
  {Kotov}},\ and\ \bibinfo {author} {\bibfnamefont {I.~M.}\ \bibnamefont
  {Mitrankov}},\ }\bibfield  {title} {\bibinfo {title} {{Production of light
  flavor hadrons in small systems measured by PHENIX at RHIC}},\ }\href
  {https://doi.org/10.1088/1402-4896/abfd64} {\bibfield  {journal} {\bibinfo
  {journal} {Phys. Scripta}\ }\textbf {\bibinfo {volume} {96}},\ \bibinfo
  {pages} {084010} (\bibinfo {year} {2021})}\BibitemShut {NoStop}%
\bibitem [{\citenamefont {Adare}\ \emph
  {et~al.}(2011{\natexlab{b}})\citenamefont {Adare} \emph {et~al.}}]{pp}%
  \BibitemOpen
  \bibfield  {author} {\bibinfo {author} {\bibfnamefont {A.}~\bibnamefont
  {Adare}} \emph {et~al.} (\bibinfo {collaboration} {PHENIX Collaboration}),\
  }\bibfield  {title} {\bibinfo {title} {{Measurement of neutral mesons in
  $p$$+$$p$ collisions at $\sqrt{s}=200$ GeV and scaling properties of hadron
  production}},\ }\href {https://doi.org/10. 1103/PhysRevD.83.052004}
  {\bibfield  {journal} {\bibinfo  {journal} {Phys. Rev. D}\ }\textbf {\bibinfo
  {volume} {83}},\ \bibinfo {pages} {052004} (\bibinfo {year}
  {2011}{\natexlab{b}})}\BibitemShut {NoStop}%
\bibitem [{\citenamefont {Adare}\ \emph {et~al.}(2016)\citenamefont {Adare}
  \emph {et~al.}}]{FlowCuAu}%
  \BibitemOpen
  \bibfield  {author} {\bibinfo {author} {\bibfnamefont {A.}~\bibnamefont
  {Adare}} \emph {et~al.} (\bibinfo {collaboration} {PHENIX Collaboration}),\
  }\bibfield  {title} {\bibinfo {title} {{Measurements of directed, elliptic,
  and triangular flow in Cu$+$Au collisions at $\sqrt{s_{_{NN}}}=200$ GeV}},\
  }\href {https://doi.org/10.1103/PhysRevC.94.054910} {\bibfield  {journal}
  {\bibinfo  {journal} {Phys. Rev. C}\ }\textbf {\bibinfo {volume} {94}},\
  \bibinfo {pages} {054910} (\bibinfo {year} {2016})}\BibitemShut {NoStop}%
\bibitem [{\citenamefont {Abelev}\ \emph {et~al.}(2007)\citenamefont {Abelev}
  \emph {et~al.}}]{Levy}%
  \BibitemOpen
  \bibfield  {author} {\bibinfo {author} {\bibfnamefont {B.}~\bibnamefont
  {Abelev}} \emph {et~al.} (\bibinfo {collaboration} {STAR Collaboration}),\
  }\bibfield  {title} {\bibinfo {title} {{Strange particle production in
  $p$$+$$p$ collisions at $\sqrt{s}=200$ GeV}},\ }\href
  {https://doi.org/10.1103/PhysRevC.75.064901} {\bibfield  {journal} {\bibinfo
  {journal} {Phys. Rev. C}\ }\textbf {\bibinfo {volume} {75}},\ \bibinfo
  {pages} {064901} (\bibinfo {year} {2007})}\BibitemShut {NoStop}%
\bibitem [{\citenamefont {Acharya}\ \emph {et~al.}(2020)\citenamefont {Acharya}
  \emph {et~al.}}]{Pi_eta_uu}%
  \BibitemOpen
  \bibfield  {author} {\bibinfo {author} {\bibfnamefont {U.}~\bibnamefont
  {Acharya}} \emph {et~al.} (\bibinfo {collaboration} {PHENIX Collaboration}),\
  }\bibfield  {title} {\bibinfo {title} {{Production of
  ${\ensuremath{\pi}}^{0}$ and $\ensuremath{\eta}$ mesons in {U}+{U} collisions
  at $\sqrt{s_{_{NN}}}=192$ GeV}},\ }\href
  {https://doi.org/10.1103/PhysRevC.102.064905} {\bibfield  {journal} {\bibinfo
   {journal} {Phys. Rev. C}\ }\textbf {\bibinfo {volume} {102}},\ \bibinfo
  {pages} {064905} (\bibinfo {year} {2020})}\BibitemShut {NoStop}%
\bibitem [{\citenamefont {Adamczyk}\ \emph {et~al.}(2016)\citenamefont
  {Adamczyk} \emph {et~al.}}]{STAR_v2_phi_AuAu}%
  \BibitemOpen
  \bibfield  {author} {\bibinfo {author} {\bibfnamefont {L.}~\bibnamefont
  {Adamczyk}} \emph {et~al.} (\bibinfo {collaboration} {STAR Collaboration}),\
  }\bibfield  {title} {\bibinfo {title} {{Centrality and Transverse Momentum
  Dependence of Elliptic Flow of Multistrange Hadrons and $\ensuremath{\phi}$
  Meson in Au$+$Au Collisions at $\sqrt{s_{_{NN}}}=200$ GeV}},\ }\href
  {https://doi.org/10.1103/PhysRevLett.116.062301} {\bibfield  {journal}
  {\bibinfo  {journal} {Phys. Rev. Lett.}\ }\textbf {\bibinfo {volume} {116}},\
  \bibinfo {pages} {062301} (\bibinfo {year} {2016})}\BibitemShut {NoStop}%
\bibitem [{\citenamefont {Huang}(2013)}]{v2_UU}%
  \BibitemOpen
  \bibfield  {author} {\bibinfo {author} {\bibfnamefont {S.}~\bibnamefont
  {Huang}},\ }\bibfield  {title} {\bibinfo {title} {{Measurements of identified
  particle anisotropic flow in Cu$+$Au and U$+$U collisions by PHENIX
  experiment}},\ }\href
  {https://doi.org/https://doi.org/10.1016/j.nuclphysa.2013.02.038} {\bibfield
  {journal} {\bibinfo  {journal} {Nucl. Phys. A}\ }\textbf {\bibinfo {volume}
  {904}},\ \bibinfo {pages} {417c} (\bibinfo {year} {2013})},\ \bibinfo {note}
  {the Quark Matter 2012}\BibitemShut {NoStop}%
\bibitem [{\citenamefont {Lin}\ and\ \citenamefont {Ko}(2002)}]{StringMelting}%
  \BibitemOpen
  \bibfield  {author} {\bibinfo {author} {\bibfnamefont {Z.-W.}\ \bibnamefont
  {Lin}}\ and\ \bibinfo {author} {\bibfnamefont {C.~M.}\ \bibnamefont {Ko}},\
  }\bibfield  {title} {\bibinfo {title} {{Partonic effects on the elliptic flow
  at relativistic heavy ion collisions}},\ }\href
  {https://doi.org/10.1103/PhysRevC.65.034904} {\bibfield  {journal} {\bibinfo
  {journal} {Phys. Rev. C}\ }\textbf {\bibinfo {volume} {65}},\ \bibinfo
  {pages} {034904} (\bibinfo {year} {2002})}\BibitemShut {NoStop}%
\bibitem [{\citenamefont {Sj{\"o}strand}\ \emph {et~al.}(2015)\citenamefont
  {Sj{\"o}strand}, \citenamefont {Ask}, \citenamefont {Christiansen},
  \citenamefont {Corke}, \citenamefont {Desai}, \citenamefont {Ilten},
  \citenamefont {Mrenna}, \citenamefont {Prestel}, \citenamefont {Rasmussen},\
  and\ \citenamefont {Skands}}]{PYTHIA8}%
  \BibitemOpen
  \bibfield  {author} {\bibinfo {author} {\bibfnamefont {T.}~\bibnamefont
  {Sj{\"o}strand}}, \bibinfo {author} {\bibfnamefont {S.}~\bibnamefont {Ask}},
  \bibinfo {author} {\bibfnamefont {J.~R.}\ \bibnamefont {Christiansen}},
  \bibinfo {author} {\bibfnamefont {R.}~\bibnamefont {Corke}}, \bibinfo
  {author} {\bibfnamefont {N.}~\bibnamefont {Desai}}, \bibinfo {author}
  {\bibfnamefont {P.}~\bibnamefont {Ilten}}, \bibinfo {author} {\bibfnamefont
  {S.}~\bibnamefont {Mrenna}}, \bibinfo {author} {\bibfnamefont
  {S.}~\bibnamefont {Prestel}}, \bibinfo {author} {\bibfnamefont {C.~O.}\
  \bibnamefont {Rasmussen}},\ and\ \bibinfo {author} {\bibfnamefont {P.~Z.}\
  \bibnamefont {Skands}},\ }\bibfield  {title} {\bibinfo {title} {{An
  introduction to {PYTHIA} 8.2}},\ }\href
  {https://doi.org/10.1016/j.cpc.2015.01.024} {\bibfield  {journal} {\bibinfo
  {journal} {Comp. Phys. Commun.}\ }\textbf {\bibinfo {volume} {191}},\
  \bibinfo {pages} {159} (\bibinfo {year} {2015})}\BibitemShut {NoStop}%
\bibitem [{\citenamefont {Alver}\ \emph {et~al.}(2011)\citenamefont {Alver}
  \emph {et~al.}}]{dNch_deta}%
  \BibitemOpen
  \bibfield  {author} {\bibinfo {author} {\bibfnamefont {B.}~\bibnamefont
  {Alver}} \emph {et~al.},\ }\bibfield  {title} {\bibinfo {title}
  {{Charged-particle multiplicity and pseudorapidity distributions measured
  with the PHOBOS detector in Au$+$Au, Cu$+$Cu, $d$$+$Au, and $p$$+$$p$
  collisions at ultrarelativistic energies}},\ }\href
  {https://doi.org/10.1103/PhysRevC.83.024913} {\bibfield  {journal} {\bibinfo
  {journal} {Phys. Rev. C}\ }\textbf {\bibinfo {volume} {83}},\ \bibinfo
  {pages} {024913} (\bibinfo {year} {2011})}\BibitemShut {NoStop}%
\bibitem [{\citenamefont {Demortier}(2008)}]{p_val}%
  \BibitemOpen
  \bibfield  {author} {\bibinfo {author} {\bibfnamefont {L.}~\bibnamefont
  {Demortier}},\ }\bibfield  {title} {\bibinfo {title} {{P Values and Nuisance
  Parameters}},\ }in\ \href@noop {} {\emph {\bibinfo {booktitle} {PHYSTAT-LHC
  Workshop on Statistical Issues for LHC Phys.}}}\ (\bibinfo {year} {2008})\
  p.~\bibinfo {pages} {23},\ \bibinfo {note}
  {http://cds.cern.ch/record/1099967}\BibitemShut {NoStop}%
\bibitem [{\citenamefont {Dusling}\ and\ \citenamefont
  {Venugopalan}(2012)}]{RecombinationFlow1}%
  \BibitemOpen
  \bibfield  {author} {\bibinfo {author} {\bibfnamefont {K.}~\bibnamefont
  {Dusling}}\ and\ \bibinfo {author} {\bibfnamefont {R.}~\bibnamefont
  {Venugopalan}},\ }\bibfield  {title} {\bibinfo {title} {{Azimuthal
  Collimation of Long Range Rapidity Correlations by Strong Color Fields in
  High Multiplicity Hadron-Hadron Collisions}},\ }\href
  {https://doi.org/10.1103/physrevlett.108.262001} {\bibfield  {journal}
  {\bibinfo  {journal} {Phys. Rev. Lett.}\ }\textbf {\bibinfo {volume} {108}},\
  \bibinfo {pages} {262001} (\bibinfo {year} {2012})}\BibitemShut {NoStop}%
\bibitem [{\citenamefont {Ortiz~Velasquez}\ \emph {et~al.}(2013)\citenamefont
  {Ortiz~Velasquez}, \citenamefont {Christiansen}, \citenamefont
  {Cuautle~Flores}, \citenamefont {Maldonado~Cervantes},\ and\ \citenamefont
  {Paic}}]{RecombinationFlow2}%
  \BibitemOpen
  \bibfield  {author} {\bibinfo {author} {\bibfnamefont {A.}~\bibnamefont
  {Ortiz~Velasquez}}, \bibinfo {author} {\bibfnamefont {P.}~\bibnamefont
  {Christiansen}}, \bibinfo {author} {\bibfnamefont {E.}~\bibnamefont
  {Cuautle~Flores}}, \bibinfo {author} {\bibfnamefont {I.~A.}\ \bibnamefont
  {Maldonado~Cervantes}},\ and\ \bibinfo {author} {\bibfnamefont
  {G.}~\bibnamefont {Paic}},\ }\bibfield  {title} {\bibinfo {title} {{Color
  Reconnection and Flow like Patterns in $p$$+$$p$ Collisions}},\ }\href
  {https://doi.org/10.1103/physrevlett.111.042001} {\bibfield  {journal}
  {\bibinfo  {journal} {Phys. Rev. Lett.}\ }\textbf {\bibinfo {volume} {111}},\
  \bibinfo {pages} {042001} (\bibinfo {year} {2013})}\BibitemShut {NoStop}%
\bibitem [{\citenamefont {Sibirtsev}\ \emph {et~al.}(2006)\citenamefont
  {Sibirtsev}, \citenamefont {Hammer}, \citenamefont {Meissner},\ and\
  \citenamefont {Thomas}}]{Phi_cross_section}%
  \BibitemOpen
  \bibfield  {author} {\bibinfo {author} {\bibfnamefont {A.}~\bibnamefont
  {Sibirtsev}}, \bibinfo {author} {\bibfnamefont {H.-W.}\ \bibnamefont
  {Hammer}}, \bibinfo {author} {\bibfnamefont {U.-G.}\ \bibnamefont
  {Meissner}},\ and\ \bibinfo {author} {\bibfnamefont {A.}~\bibnamefont
  {Thomas}},\ }\bibfield  {title} {\bibinfo {title} {{phi-meson photoproduction
  from nuclei}},\ }\href {https://doi.org/10. 1140/epja/i2006-10070-4}
  {\bibfield  {journal} {\bibinfo  {journal} {Eur. Phys. J. A}\ }\textbf
  {\bibinfo {volume} {29}},\ \bibinfo {pages} {209} (\bibinfo {year}
  {2006})}\BibitemShut {NoStop}%
\end{thebibliography}

%
 
\end{document}